\documentclass[11pt,superscriptaddress]{revtex4-2}
\usepackage{graphicx}
\usepackage{amssymb}
\usepackage{amsfonts}
\usepackage{amsbsy}
\usepackage{amsmath}
\usepackage{textcomp}
\usepackage{amsmath}
\usepackage{amssymb}
\usepackage{multirow} 
\usepackage{braket}
\usepackage{subfig}
\usepackage{siunitx}
\usepackage{mathtools}

\newcommand{\pdd}[1]{\frac{\partial}{\partial#1}}

\begin{document}
\title{Spectral Densities, Structured Noise and Ensemble Averaging  within Open Quantum Dynamics}
\author{Yannick Marcel Holtkamp} 
\affiliation{School of Science, Constructor University,
	Campus Ring 1, 28759 Bremen, Germany}
\author{Emiliano Godinez-Ramirez} 
\altaffiliation{Present address: Department of Physics, Technical University of Munich, 
        James-Franck-Straße 1, 85748 Garching, Germany }
\affiliation{School of Science, Constructor University,
	Campus Ring 1, 28759 Bremen, Germany}
\author{Ulrich Kleinekath\"ofer}
\email{ukleinekathoefer@constructor.university}
\affiliation{School of Science, Constructor University,
	Campus Ring 1, 28759 Bremen, Germany}

\begin{abstract}
  Although recent advances in simulating open quantum systems have lead to significant progress, the applicability of numerically exact methods is still restricted to rather small systems.  Hence, more approximate methods remain relevant due to their computational efficiency, enabling simulations of larger systems over extended timescales. In this study, we present advances for one such method, namely the Numerical Integration of Schrödinger Equation (NISE). Firstly, we introduce a modified ensemble-averaging procedure that improves the long-time behavior of the thermalized variant of the NISE scheme, termed Thermalized NISE. Secondly, we demonstrate how to use the NISE in conjunction with (highly) structured spectral densities by utilizing a noise generating algorithm for arbitrary structured noise. This algorithm also serves as a tool for establishing best practices in determining spectral densities from excited state calculations along molecular dynamics or quantum mechanics/molecular mechanics trajectories. Finally, we assess the ability of the NISE approach to calculate absorption spectra and
  demonstrate the utility of the proposed modifications by determining population dynamics. 
\end{abstract}

\maketitle

\section{Introduction}

Isolated quantum systems are excellent models for understanding the basic principles of quantum physics. In practice, however, quantum systems are never truly isolated. Phenomena such as charge and energy transfer, relaxation processes and quantum coherences in
man-made quantum devices \cite{koch16a,hu20a} or biological systems such as photosynthetic complexes \cite{cao20a,mait23a} can only be explained in detail within the framework of open quantum systems, where the systems are separated into a primary system and a bath. 
Solving such quantum dynamical problems analytically is only possible for very simple setups, and usually also computationally (very) challenging for larger systems. 
The most straightforward way to treat such a system would be to describe the system as well as all the bath degrees of freedom quantum dynamically. This can be efficiently done with variational methods, e.g., using the MCTDH approach \cite{wort08a,meye12a}. Such a treatment is restricted to a relatively low number of degrees of freedom due to the large computational effort. However, usually the exact time evolution of the bath degrees of freedom is not of interest. This led to the Multi-Layer MCTDH (ML-MCTDH) approach \cite{vend11a, wang15c}, where bath degrees of freedom are combined, but the simulation of the system remains exact. This scheme has been, e.g., used to simulate the quantum dynamics of the Fenna-Matthews-Olson (FMO) complex with 8 sites and 592 bath degrees of freedom \cite{schu16a}. Another numerically exact variational method is the hierarchy of Davydov Ansätze \cite{chen18f,zhao23a}, which can, depending on the problem, be numerically less costly than the ML-MCTDH scheme. A more recent development for treating the bath explicitly is the Automated Compression of Environments method \cite{cygo22a}, which uses automatic tensor contraction to automatically reduce the computational effort for the bath degrees of freedom. A general issue that arises with the full explicit description of the bath degrees of freedom is that, due to the time-reversible nature of the Schrödinger equation, a true thermal equilibrium state can only be reached in the limit of infinite bath degrees of freedom. However, a practical equilibrium state for the system degrees of freedom can be achieved with a finite number of bath degrees of freedom. The number of bath degrees of freedom needed to achieve such an effective thermal equilibrium state increases with the simulation time, so that the computational method for explicit bath descriptions increases not only with the system size but also more than linearly with the simulation time.

To circumvent such issues, methods have been developed that can handle baths with effectively infinite degrees of freedom. Instead of treating the bath degrees of freedom explicitly, these methods assume a harmonic oscillator bath and linear coupling to the system degrees of freedom described by a spectral density \cite{nitz06a,may11a}. Among those schemes are the QUAsiadiabatic Propagator Path Integral (QUAPI) method \cite{makr95a,thor00a} and the popular Hierarchical Equations of Motions (HEOM) formalism \cite{tani20a,iked20a} as well as the Hierarchy of Pure States (HOPS) \cite{sues14a,hart17a,varv21a,gera23a} method. While reducing the computational efforts compared to methods that treat the bath explicitly, these methods still remain challenging for very large systems and complicated spectral densities.
The original HEOM formalism \cite{tani89}, for instance, is limited to simple Drude-Lorentz spectral densities due to its reliance on  exponentially decaying bath correlation functions. Its application is confined to small systems and spectral densities with only a few peaks due to an unfavorable computational scaling. Realistic systems, however, tend to be more complicated, prompting the development of several HEOM extensions. On the computational side, efficient GPU implementations such as the GPU-HEOM implementation  \cite{krei13b}, and, more recently, tensor train representations \cite{shi18a, borr21a,mang23a} have been used to handle the numerical demands. Several advancements have also been made by developing HEOM variants that allow for non-exponential terms in the bath correlation functions \cite{tang15a, iked20a, chen24a} and finding efficient ways of decomposing arbitrary spectral densities into such terms \cite{chen22c, xu22a, taka24a}.
With these advancements, the HEOM approach has, for example, recently been applied to a system with 96 chlorophyll  molecules, each with a 148-mode decomposition of an experimental FMO correlation function \cite{shi21a}. Another approach, that is closely related to the HEOM approach, the HOPS scheme\cite{sues14a,hart17a}. It includes non-Markovian effects via stochastic bath realizations. Its adaptive basis variant, ad-Hops \cite{varv21a,gera23a}, promises good system size scalability; however, the applications so far have been limited to few bath modes. At the same time, some progress is being made recently with new approaches which are, however, mostly limited to small systems or a few bath modes \cite{plea20a, bose22a}.  Within the path integral-based methods, the Modular Path Integral formalism \cite{makr18a,kund22a} is worth mentioning, as it allows for an improved scaling of the computational costs if the Hamiltonian can be dissected into identical subunits. 

Even with this progress in the field, more approximate methods, such as Redfield theory \cite{ishi09c} or ensemble-averaged wave packet methods \cite{jans04a,zhu08c,mego11a}  remain relevant, as they allow for computationally more efficient calculations and, thereby, allowing for quantum dynamics calculations of larger systems on longer time scales. 
One such method, the Numerical Integration of Schrödinger Equation (NISE) \cite{jans06a,jans18a,bond20a,holt23a},  also known as Ehrenfest scheme without back-reaction \cite{aght12a}, has, e.g., recently been applied to chlorosomes with over 27,000 individual pigments \cite{eric23a,eric23b,eric24a}. Systems of such scale remain out of reach for numerically exact methods, especially if non-Gaussian baths, which NISE can treat \cite{roy11a}, are of relevance. Indications for  non-Gaussian contributions have been found, e.g., in the light harvesting complex 2 (LH2) from purple bacteria \cite{mour01a,sing19a}. An overview of non-Gaussian effects can be found in Ref.~\onlinecite{fara17a}. Furthermore, NISE can treat non-Condon effects for absorption calculations absent from most other methods, although a HEOM variant for non-Condon effects has been developed \cite{seib18a}. The NISE approach will be at the main focus of this study.

The NISE scheme is a mixed quantum-classical method, i.e., it only treats the system quantum mechanically, while the bath is treated classically. Furthermore, the back reaction of the system onto the bath is neglected, leading to one of the greatest advantages of the NISE, i.e.,  the bath becomes completely independent of the system and can be calculated prior to the quantum dynamical calculations. The interactions between system and bath are then included as fluctuating site energies in effective Hamiltonian operators. Previously, these fluctuating site energies have been obtained either from excited state calculations along molecular dynamics (MD) or quantum mechanics/molecular mechanics (QM/MM) trajectories \cite{damj02a,mait20a,cign22a,mait23a,eric23a,sarn24a} or generated following simple analytic spectral densities \cite{aght12a,fuji12a,jans18a,holt23a}. Similar to the progress made for the HEOM approach, it is also of interest to enable the use of the NISE scheme for complicated spectral densities. To this end, we review and test a noise-generating algorithm that can easily generate site energy fluctuations corresponding to (very) complicated spectral densities. This “noise” can then be used in a NISE calculation for cases in which the spectral density is known. 
In the present study, we utilize this algorithm for an additional purpose. Spectral densities in our and several other groups are determined based on excited state calculations along MD or QM/MM trajectories \cite{damj02a,olbr10a,vall12a,lee16a,kim18a,loco18b,mait20a,cign22a,mait23a,sarn24a}. These calculations are numerically expensive and questions arise concerning the simulation time step, length of trajectory or damping function to obtain the most accurate spectral density based on the MD or QM/MM calculations. Being able to generate long artificial noise trajectories based on an arbitrarily complicated spectral densities makes it possible to test the necessary time step, length of trajectory or damping function since the spectral density which should be reproduced is known already.   

As mentioned above, the NISE approach is an approximate method and, as such, has its shortcomings. One approximation of the NISE scheme entails the aforementioned neglect of the effect of the quantum system on its classical thermal environment, i.e., the thermal environment affects the quantum system but not vice versa, hence, the alternative name Ehrenfest scheme without back-reaction.  This approximation leads to the first shortcoming of the method. Large system-bath couplings can lead to strong perturbations in the thermal surrounding. Since these perturbations are neglected, the NISE formalism is only applicable to small or medium system-bath coupling strengths. The assumption of a classical bath introduces another approximation. This assumption implicitly includes a high-temperature approximation so that detailed balance is not satisfied at long simulation times and, consequently, the correct Boltzmann distributions are not produced in the long-term limit \cite{para06a}. 

The NISE is frequently employed when more precise methods are computationally unfeasible due to the system size or a bath that cannot be accurately represented by computationally efficient methods due to their typically presumptions of a harmonic bath linearly coupled to the system. Depending on the method, the latter can be due to an anharmonic bath, a non-linear coupling to the bath, or the necessity to consider non-Condon effects. Additionally, the drawbacks of the NISE approach have to be acceptable, i.e., the NISE scheme only yields reasonable results for systems with low to medium system-bath couplings and  when the long-term dynamics and equilibrium populations are of less interest. The latter is, for example,  the case for linear absorption spectra.

 Regarding the wrong long-term dynamics, several corrections have been attempted to fix this shortcoming \cite{bast06c,aght12a,jans18a,holt23a}, but the correct thermal distribution at long times was only achieved approximately in all of these cases. In a previous study \cite{holt23a}, we realized that the averaging procedure of the individual realizations in the ensemble-averaged procedure is a root cause of this discrepancy due to the non-linear exponential function in the Boltzmann distribution. In the present study, we propose a new averaging procedure that addresses this issue effectively. By improving the long-term dynamics, we hope to make the NISE approach more useful for issues like identifying exciton pathways in large photosynthetic structures.

As mentioned previously, the NISE scheme is often used for linear absorption, as it is an application where the drawbacks are less severe. The noise-generating algorithm made us want to test it against some other popular linear absorption methods. As benchmark, we again use the HEOM formalism and, in addition, test the  Redfield theory \cite{novo10a,reng13a,juri15b} and the Full Cumulant Expansion (FCE) scheme \cite{cupe20b} which are also often employed for multichromophoric systems that we have in mind as potential application of the proposed improvements.

The article is structured as follows: Section~\ref{sec:NISE} reviews the foundation of the NISE approach and its limitations, while in Section~\ref{sec:new_averaging_procedure}, a modified ensemble-averaging procedure is presented which is designed to achieve a proper 
long-time dynamics within the in the NISE framework.
 The noise-generation algorithm is detailed in Section~\ref{sec:NoiseGen} together with its ability to generate noise that follows arbitrarily complicated spectral densities. In Section~\ref{sec:SD-reconstruction}, we utilize the noise generating algorithm to establish a best practice for calculating spectral densities. 
 An example for population dynamics using the NISE scheme and its variants for systems with highly structured  spectral densities is shown in Section~\ref{sec:population_dynamics}, while in Section\ref{sec:absorption}, the accuracy of linear absorption spectra based on the NISE scheme is assessed. The article concludes with a summary of our findings and potential future research directions in Section~\ref{sec:Conclusions}.

\section{Numerical Integration of the Schrödinger Equation}
\label{sec:NISE}
\subsection{System-Bath Approach}

To model quantum effects in large disordered systems, one usually uses one of
the many variants of system-bath approaches. While the NISE can be derived for quite general system-bath approaches \cite{jans06a,bond20a}, we will limit our derivation to a linearly coupled harmonic oscillator bath to enable comparisons with common density matrix approaches. In these schemes, the Hamiltonian
$\hat{H}$ is split into a (primary) system part $\hat{H}_{S}$ and a bath part
$\hat{H}_{B}$
\begin{equation}
\hat{H}=\hat{H}_{S}+\hat{H}_{B} +\hat{H}_{SB}~.
\end{equation}
As system Hamiltonian, a standard tight-binding model is employed with site energies $E_n$ at site $n$ and inter-site couplings $V_{mn}$
\begin{equation}
	\hat{H}_S=\sum_{n=1}^N E_n \ket{n} \bra{n} +\sum_{m=1}^N\sum_{n=1}^N
	V_{mn} \ket{m} \bra{n}.
	\label{H_S}
\end{equation} 
In principle, the site energies and couplings can be explicit time-dependent, for instance, due to an externally applied electric field. However, for simplicity and to avoid confusion with time dependencies arising from interactions with the environment, the site energies, couplings, and consequently the system Hamiltonian are considered to be time-independent in this work.
While the bath is assumed to consist of harmonic oscillators,  the system-bath
coupling Hamiltonian $\hat{H}_{SB}$ is given as a sum of products of system and
bath operators $\hat{\Phi}_j$. In the present study we adopt a site-local form of the system
part of the system-bath interaction \cite{may11a, aght12a}, i.e.,
\begin{equation}
 \hat{H}_{SB}=\sum_n \hat{\Phi}_n \ket{n}\bra{n}~,
\label{hsb}
\end{equation}
with the sum over the sites $n$. Similar results as those shown below can,
however, also be obtained for a more general form of  $\hat{H}_{SB}$ while in
the present study we have exciton transfer systems in mind. Within the harmonic bath and linear coupling approximations, the system-bath interaction is fully characterized using the spectral density  $J_n(\omega)$ at site $n$ \cite{damj02a,guti10a,may11a,aght12a,vall12a}
\begin{equation}
	J_n(\omega)=\frac{\beta\omega}{\pi} \int\limits_{0}^{\infty} dt C_n(t)\cos(\omega t)
	=\frac{\beta\omega}{2\pi} \tilde{C}_n(\omega)
	\label{eq:jw}
\end{equation}
with $\beta=1/(k_B T)$ denoting the inverse temperature. The main ingredient in this
expression is the energy gap autocorrelation function $C_n(t)=\braket{ \Delta
	E_n(t)\Delta E_n(0) }$. Moreover, we have used the relation $C_n(t)=C_n(-t)$ for real autocorrelation functions and the definition of the Fourier transformation in the form 
$\tilde{C}(\omega)$=$\int_{-\infty}^{\infty} C(t)\exp(-i\omega t) dt $. Please note that this form of the spectral density is connected to the one in the  Caldeira-Leggett  model \cite{cald83a} by $J_{CL,n}(\omega)=\frac{\pi}{\hbar}J_{n}(\omega)$.


\subsection{Ensemble-Averaged Wave Packet dynamics}

As an alternative to density matrix simulations, one can use
an ensemble-averaged wave packet approach termed the NISE approach. 
Let us take 
 a look at the time-dependent Schrödinger equation for the complete
system including the full Hamiltonian
\begin{equation}
  i\hbar\pdd{t}\ket{\Psi(t)}= \hat{H} \ket{\Psi(t)}=(\hat{H}_S + \hat{H}_B +
\hat{H}_{SB})\ket{\Psi(t)}. \label{schrod}
\end{equation}
Since the bath part is usually very high-dimensional, a direct solution of this equation is infeasible in almost all cases. One can now assume that the bath  stays in equilibrium if initially in
equilibrium, which is a standard assumption in perturbation theories \cite{may11a}. Based on this simplification, it is straightforward to derive an 
 an effective Hamiltonian for the relevant system \cite{aght12a}
\begin{eqnarray}
\hat{H}_S^{\rm eff}(t)&=&\sum_m (E_m + \Delta E_m(t)) \ket{m}\bra{m} + \sum_{n \neq m}
V_{nm} \ket{n}\bra{m}~.
\label{eq:heff}
\end{eqnarray}
The term $\Delta E_m(t)=\bra{\Psi_B(t)}\hat{\Phi}_m\ket{\Psi_B(t)}$ represents
the site energy fluctuations resulting from the coupling between system and
environment. When performing a proper ensemble average, solving the time-dependent Schrödinger equation with this
effective Hamiltonian yields the dynamics of the relevant system in the case
of weak system-bath coupling. 
We want to emphasize that the NISE scheme in its general form can both handle an anharmonic/non-Gaussian bath and account for non-Condon effects, i.e., the restriction mentioned above for most density matrix approaches do not apply \cite{jans06a,bond20a}.  

The expectation value of a system operator can then be written as 
 \begin{equation}
 	\braket{A}=
 	\frac{1}{N_\alpha}\sum_{\alpha=1}^{N_\alpha} 
 	\braket{\Psi_S^\alpha (t)|A|  \Psi_S^\alpha (t)}  =\overline{  \braket{\Psi^\alpha_S(t)|A|  \Psi^\alpha_S(t)}}=\mbox{tr}(\rho^{\text{NISE}}A)
 \end{equation}
 with $N_\alpha$ being the number of samples and $\rho^{\text{NISE}}=\overline{\ket{\Psi^\alpha_S(t)}\bra{\Psi^\alpha_S(t)}}$.
In site basis, the density matrix is thus given by 
\begin{equation}
\rho_{mn}^{\text{NISE}} (t)  = \bra{m } \overline{  \Psi_S^\alpha (t) \rangle \langle \Psi_S^\alpha (t)}  \ket{n}   
= \bra{m} \rho^{\text{NISE}} \ket{n}~.
\label{eq:NISE_density_Matrix}
\end{equation}
The aim of the present study is to generate fluctuating site  energies which
lead to predefined spectral densities and to subsequently compare the results of
density matrix and ensemble-averaged wave packet dynamics. For this comparison,
one will have to keep in mind that by solving the ensemble-averaged wave packet
dynamics, one implicitly introduces a high-temperature limit. Thus, detailed
balance cannot be obtained without additional correction factors, a topic which 
will be discussed in more detail below.

\subsection{Thermal Correction in Ensemble-Averaged Dynamics}
\label{subsec:thermal_correction_ensemble_averaged_dynamics}

To lay the groundwork for introducing the thermal correction, a specific method for the propagation of the wave function within ensemble averaged dynamics has to be established. To this end, we follow the Thermalized NISE (TNISE) procedure as detailed in Ref.~\onlinecite{jans18a}, which involves the iterative propagation of the wave function (in site basis) in small time steps
\begin{equation}\label{eq:timeevolution_revisited}
\ket{\psi(t+\Delta t)}= \hat{W}^\dagger (t) \hat{\Tilde{U}}(t+ \Delta t , t) \hat{W}(t) \ket{\psi(t)}~.
\end{equation}
From here on, we use the notation of an overhead tilde to denote that the operator or wave function has been transformed into the eigenbasis of the instantaneous Hamiltonian,  and $\hat{W}(t)$ denotes the operator representing this transformation. This transformation is applied because the time evolution operator becomes much simpler in the eigenbasis with the eigenstates $\ket{\alpha}$, i.e.,
$\hat{\Tilde{U}}(t+ \Delta t , t)= \sum_\alpha \exp (-i \epsilon_\alpha (t) \Delta t/\hbar) \ket{\alpha} \bra{\alpha}$. To introduce the thermal correction, the iterative propagation is then rewritten in the eigenbasis
\begin{equation}
\ket{\tilde{\psi}(t+\Delta t)}= \hat{S}(t) \hat{\Tilde{U}}(t+ \Delta t , t) \ket{\tilde{\psi}(t)}~,
\end{equation}
where $\hat{S}(t)= \hat{W}(t+\Delta t) \hat{W}^\dagger (t)$ represents the non-adiabatic coupling operator. 

As mentioned above, with the high-temperature approximation inherent in ensemble-averaged wave packet dynamics, it becomes necessary to introduce a thermal correction to achieve detailed balance. Following the methodology outlined in Ref.~\onlinecite{jans18a},  we incorporate a thermal correction into the off-diagonal elements of the non-adiabatic coupling matrix $S(t)$. This correction aims at accelerating the transfer from higher to lower energetic eigenstates and to slow down the transfer in the opposite direction. The thermal correction term, as proposed by Jansen, is given by \onlinecite{jans18a}
\begin{equation}\label{eq:JansenFactor_revisited}
S^{T}_{\alpha,\beta}(t)=S_{\alpha,\beta}(t) \cdot \left( (1-\delta_{\alpha,\beta})\exp\left(\frac{\Delta\epsilon_{\alpha,\beta}(t)}{4 k_B T}\right)+\delta_{\alpha,\beta} \right)~,
\end{equation}
where $S^{T}(t)$ denotes the thermalized version of $S(t)$, and $\Delta\epsilon_{\alpha,\beta}(t)$  the energy difference between eigenstates $\alpha$ and $\beta$. Using the Boltzmann factor, this term intuitively aligns better with physical principles, though it remains an ad-hoc choice. Instead of this ad-hoc choice, Ref.~\onlinecite{holt23a} introduced a machine-learned correction term, replacing the ad hoc term $\exp\left(\Delta\epsilon_{\alpha,\beta}(t) / ({4 k_B T})\right)$ with a neural network. To differentiate the method using the neural network from the TNISE scheme, it has been named Machine Learned NISE (MLNISE).

\section{Improved Averaging Procedure for Ensemble-Averaged Wave Packet Dynamics}
\label{sec:new_averaging_procedure}

\subsection{Introduction of a Modified Ensemble-Averaging Procedure}
For a time-independent Hamiltonian $\hat{H}_S$, the corresponding equilibrium density operator is given by the Boltzmann-Gibbs distribution
\begin{equation}
\hat{\rho}^{eq}(\hat{H}_S) = \frac{e^{-\beta \hat{H}_S}}{\text{Tr}(e^{-\beta \hat{H}_S})}.
\label{eq:b_eq}
\end{equation}
In the weak system-bath limit, this expression can be seen as a good approximation also when the system is coupled to an environment  \cite{geva00, gelz20a} (see also Fig.~\ref{fig:new_averaging_FMO} for the accuracy in FMO). Since in the NISE schemes, a weak system-bath coupling is assumed anyway, we are not going beyond this expression for the equilibrium density matrix. Furthermore, we do not treat any static disorder effects as they are also not included in the HEOM calculations, which we use as reference. 
In the NISE approach, the time-independent Hamiltonian corresponds to the ensemble- and time-averaged instantaneous Hamiltonian $\hat{H}_S =  \left\langle \overline{\hat{H}_S^{\rm  {eff},\alpha}(t)}\right\rangle$, where the brackets correspond to the average over time and the overline to the average over the realizations $\alpha$. 
Since in the NISE schemes the back reaction of the system dynamics on the bath is neglected, one can perform the dynamics of the different realizations independently followed by an averaging over the dynamics of the individual realizations. This procedure explains the observations in a previous study \cite{holt23a}, that the TNISE and MLNISE schemes seem to converge on average to equilibrium distributions corresponding to the respective instantaneous Hamiltonian  $\hat{H}_S^{\rm  {eff},\alpha}(t)$. We note that no formal equilibrium operator has been derived for TNISE and MLNISE and since the thermal correction is included in an ad-hoc way, it is not clear if this can be formally done. However, in the mentioned study \cite{holt23a}, the thermal distribution of TNISE and MLNIE has been observed to be close to the left-hand side of the following equation in many cases (see also Fig.~\ref{fig:new_averaging_FMO} for FMO)
\begin{equation}
\lim_{{t}\to\infty} \hat{\rho}^{\text{TNISE}} (t) \approx \left\langle \overline{\hat{\rho}^{eq}\left(\hat{H}_S^{\rm  {eff},\alpha}(t)\right)} \right\rangle \neq \hat{\rho}^{eq} \left ( \left\langle\overline{\hat{H}_S^{\rm  {eff},\alpha}(t)} \right\rangle\right ).
\label{eq:averagingissue}
\end{equation}
The left-hand side of this inequality is what one gets, while the right-hand side is what is wanted for a proper long-time limit (for weak to medium system-bath coupling). These two averages are not the same due to the non-linearity of the exponential function in Eq.~\ref{eq:b_eq}. The proper limit, at least in the case of weak system-bath coupling, is given by the right-hand side of this inequality (see also Fig.~\ref{fig:new_averaging_FMO} for these limits in the FMO complex). The observation, that TNISE and MLNISE reach a thermal equilibrium close to the left-hand side of this inequality, is the basis of the following averaging method.

To address the non-linearity, we propose a method that involves constructing “artificial'' Hamilton operators and then averaging those to create a new constructed density matrix $\hat{\rho}^{c} (t)$
\begin{equation}
\hat{\rho}^{c} (t) = \hat{\rho}^{eq}\left(  \overline{ \hat{H}_A^{\alpha}(t)}\right)~.
\end{equation}
In this way, the averaging mimics the averaging in the proper thermal limit more closely, so that the constructed density matrix $\hat{\rho}^{c} (t)$ can hopefully reach an improved equilibrium population compared to the traditionally averaged $\hat{\rho}^{\text{TNISE}} (t)$. This scheme will be tested numerically below.
For this purpose, we propose to construct these artificial time-dependent Hamilton operators $\hat{H}_A^\alpha(t)$  for every realization $\alpha$ as
\begin{equation}
\hat{H}_A^\alpha(t) = -\frac{1}{\beta}\ln(\hat{\rho}^{\text{TNISE},\alpha}(t))
= -\frac{1}{\beta}\ln(
\ket{\Psi^\alpha_S(t)}\bra{\Psi^\alpha_S(t)})~,
\label{eq:Artifitial_Hamiltonian}
\end{equation}
so that for an individual realization $\alpha$ the equilibrium matrix of the artificial Hamiltonian $\hat{H}_A^\alpha(t)$ equals the TNISE density matrix  $\hat{\rho}^{\text{TNISE},\alpha} (t)$. We need to note that in the current work the natural logarithm of an operator $\hat{A}$ is defined as 
 $	\ln(\hat{A}) = \hat{W}  \ln(\hat{\Lambda})  \hat{W}^\dagger$ ,
where $\hat{W}$ denotes the unitary operator that transforms $\hat{A}$ into its eigenbasis, such that $\hat{A}=\hat{W}  \hat{\Lambda}  \hat{W}^\dagger$. The logarithm can then be applied directly to the individual eigenvalues of $\hat{\Lambda}$.
A more general definition of the natural logarithm of an operator is any solution where $\exp(\ln(\hat{A})) = \hat{A}$. The definition employed here is not the only solution that fulfills this criterion, but is best suited in the context of this study. 
As a technical implementation note, to avoid numerical issues, we set any zero eigenvalues to a small value \( \epsilon = 10^{-10} \) to ensure stability in the computations of the logarithms.
 Using this definition, one can easily verify that the equilibrium matrix of the artificial Hamiltonian indeed equals the TNISE density matrix, i.e., $\hat{\rho}^{eq}(\hat{H}_A^\alpha(t))=\hat{\rho}^{\text{TNISE},\alpha} (t)$.
Thus, after substituting $\hat{H}_A^{\alpha}(t)$ for its definition, the expression for the constructed density matrix $\hat{\rho}^{c} (t)$ simplifies to be
\begin{equation}
\hat{\rho}^{c} (t) = \hat{\rho}^{eq}\left(  \overline{ \hat{H}_A^{\alpha}(t)}\right) = \exp{ \left( \overline{\ln(\hat{\rho}^{\text{TNISE},\alpha}(t))}\right)}.
\end{equation}

\subsection{Identifying Limitations and Proposing an Interpolation Method}
The observation, on which this modified ensemble-averaging procedure is based, only holds in the long-time limit (see Eq.~\ref{eq:averagingissue}). Unfortunately, the modified procedure does not capture the initial dynamics very well compared to the original averaging method. To rectify this, we propose an interpolation scheme that transitions between the two averaging schemes based on the lifetime of each quantum state.

The lifetimes $\tau_n$ for this interpolation are determined by fitting an exponential decay to the (non-thermalized) NISE population dynamics of each state $n$ in the site basis. Here, we note that it is not necessary to run a separate NISE calculation to obtain the population dynamics for every state. The time evolution operator $U(t,0)$ taking the system from the starting time to time $t$ can be obtained directly and can then applied to all different initial states. The interpolation is governed by the weight function $w_n(t),$ defined as
\begin{equation}
 w_n(t) = 1 - e^{-\frac{t}{5\tau_n}} ,
 \label{eq:empirical_factor}
\end{equation}
where $\tau_n$  represents the characteristic lifetime of the state (obtained by an exponential fit), and the factor 5 is an empirically determined adjustment factor. This adjustment factor is used, since the interpolation should switch to the new averaging once the long-time limit is reached, and the time span of one characteristic lifetime is certainly not enough for that.
The interpolated population for state $n$ at time $t$ can be approximated as 
\begin{equation}
 \rho_{n,n}^{\text{interpolated}}(t) = w_n(t) \rho_{n,n}^{c}(t) + (1 - w_n(t)) \rho_{n,n}^{\text{TNISE}}(t)~.
\end{equation}
This population subsequently needs to be renormalized to ensure the total population remains normalized. It has to be noted, that the method in its current form, only works for the population, i.e., diagonal elements of the density matrix. For coherences, a different way of finding characteristic lifetimes would have to be devised. 

\subsection{Improved Population Dynamics Results based on the Modified Ensemble Averaging}

\begin{figure}[tb]
\captionsetup[subfloat]{position=top,singlelinecheck=false,justification=raggedright,labelformat=brace,font=bf}
\subfloat[][]{
\includegraphics[width=0.485\textwidth]{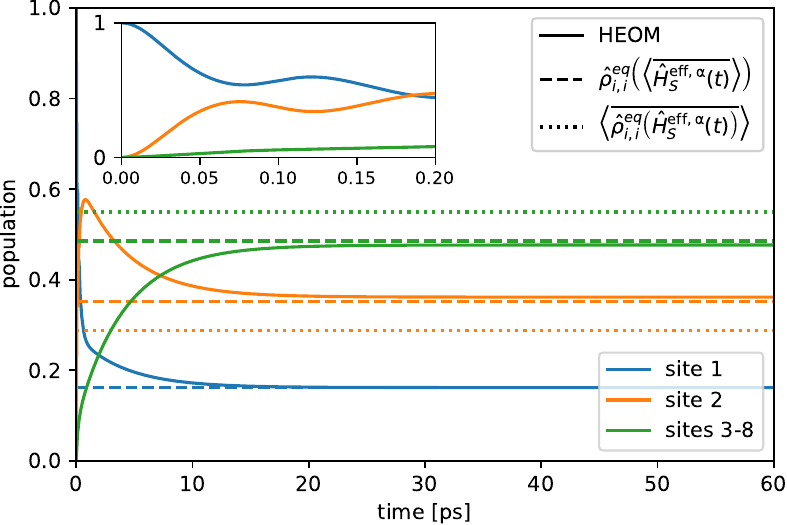}
}
\hskip0em\relax
\subfloat[][]{
\includegraphics[width=0.485\textwidth]{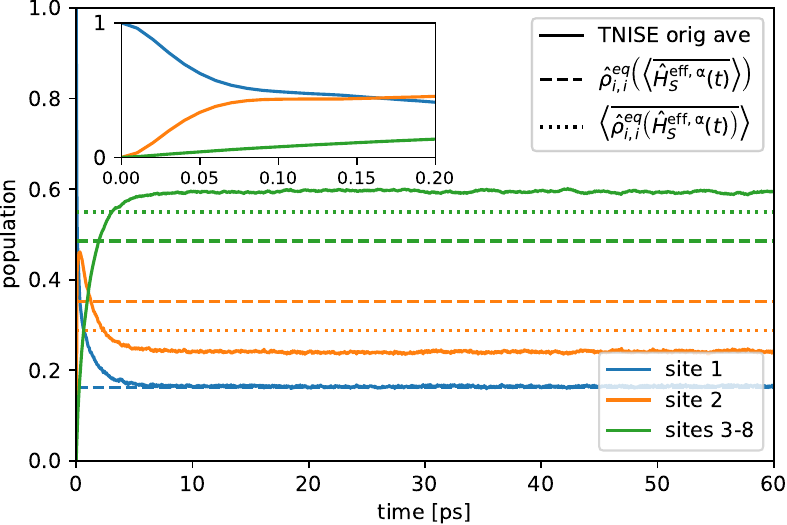}
}
\\[-4ex]
\subfloat[][]{
\includegraphics[width=0.485\textwidth]{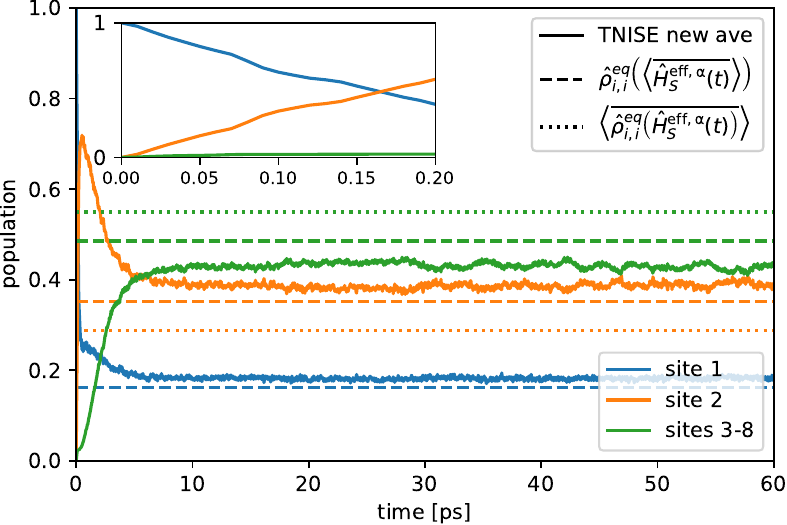}
}
\hskip0em\relax
\subfloat[][]{
\includegraphics[width=0.485\textwidth]{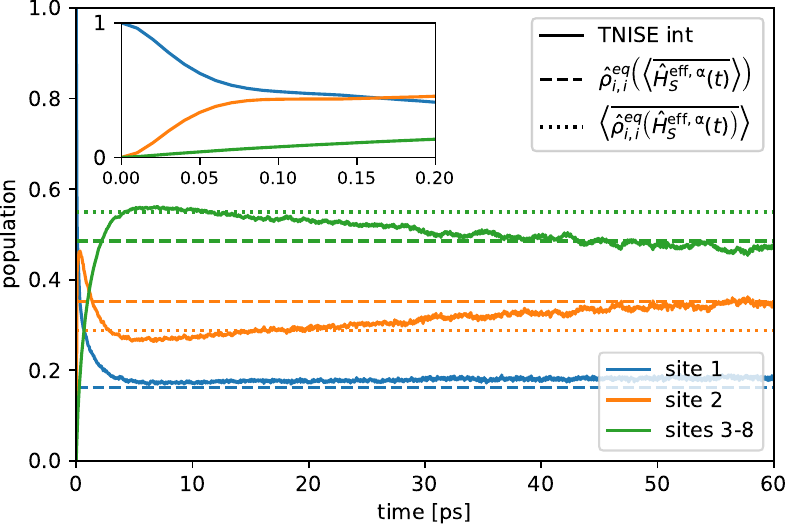}
}
\caption{Long-time population dynamics of an 8-site FMO Hamiltonian calculated with a) HEOM b) original TNISE c) new averaging TNISE without interpolation and d) new averaging TNISE with interpolation. The populations of sites 3-8 are summed together for clarity. Furthermore, the thermal distributions according to the right-hand side of inequality~\ref{eq:averagingissue} are shown as dashed lines an according to the left-hand side as dotted lines and the short-time dynamics of the first 0.2 ps are shown as an inset. 10,000 realizations were used for the different TNISE results.}
\label{fig:new_averaging_FMO}
\end{figure}

In this subsection, we explore the enhanced performance of the modified ensemble-averaging procedure by comparing results obtained using this approach with those based on the traditional averaging approach and by benchmarking it against the HEOM formalism in the case of two-site systems and a model FMO system. Initially, we explore the effects on the model FMO system. We use an 8-site FMO Hamiltonian coupled to a two-peak spectral density with a reorganization energy of $\lambda = 200 ~ \text{cm}^{-1}$. The spectral density is the two-peak spectral density introduced later in Sec. \ref{sec:population_dynamics}, scaled to the correct reorganization energy. The same model will be used in later sections, where the choice of parameters will be explained more thoroughly. For example, the absorption spectra for these parameters are calculated in Sec.~\ref{sec:absorption}. In the current section, the FMO system only serves to demonstrate the wrong thermal equilibrium of the standard averaging method, how the new method addresses it and further demonstrates that the new averaging method can also work with larger systems. Fig.~\ref{fig:new_averaging_FMO} shows the population dynamics of the FMO system until thermal equilibrium is reached, obtained by the HEOM and the TNISE methods. The MLNISE models trained in Ref.~\onlinecite{holt23a} are not suitable for this comparison, as they were only trained on much shorter timescales. In panel a, you can see that the equilibrium value is very close to the right side of the inequality \ref{eq:averagingissue}, which is normal for low system-bath coupling. The Mean Squared Error (MSE) between the final HEOM population and the right-hand side of Eq.~\ref{eq:averagingissue} is $1.4 \cdot 10^{-5}$. Panel b shows the results from the TNISE scheme with the original averaging method which reaches an equilibrium closer to the left-hand side of Eq.~\ref{eq:averagingissue}, where the final population (averaged over the last 6ps) has an MSE of $1.9 \cdot 10^{-3}$ compared to the right-hand side. Furthermore, Panel c shows the result of the new averaging method without interpolation, reaching a thermal distribution much closer to the right-hand side of Eq.~\ref{eq:averagingissue} with an MSE of $3.5 \cdot 10^{-4}$, an improvement of almost an order of magnitude. At the same time,  this figure also shows the initial dynamics, diverging from the original averaging. Finally, in panel d one can see the results for the new averaging method including interpolation, thereby rectifying this issue. The transitioning to the new averaging happens perhaps too slowly in this case, as the population has not fully converged to the new averaging after the shown 60~ps. The fitted lifetimes were between 1.3~ps and 8.1~ps in this case.

\begin{figure}[tb]
\centering
\includegraphics[width=13cm]{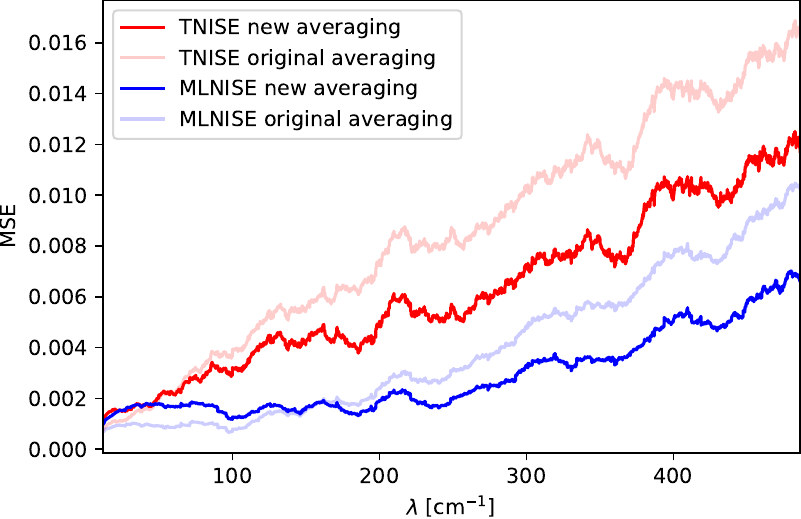}
\caption{The graph presents a running average of the mean squared error (MSE) along a 1~ps-long trajectory with 1000 realizations, compared to the HEOM results, as a function of the reorganization energy $\lambda$ for two-site systems with varying parameters. The data represents a composite of 10,000 systems, where the parameters were randomly picked from the ranges listed in Tab.~\ref{table:sampleSytemParameters}. Each point represents a moving average over 250 systems to smooth out individual discrepancies and to highlight trends more effectively.}
\label{fig:new_averaging_many_systems}
\end{figure}

\begin{figure}[tb]
\captionsetup[subfloat]{position=top,singlelinecheck=false,justification=raggedright,labelformat=brace,font=bf}
\subfloat[][]{
\includegraphics[width=0.485\textwidth]{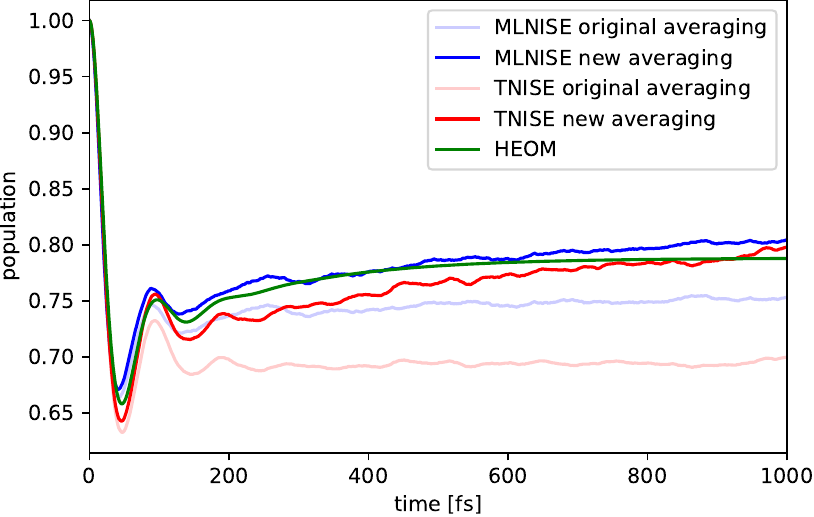}
}
\hskip0em\relax
\subfloat[][]{
\includegraphics[width=0.485\textwidth]{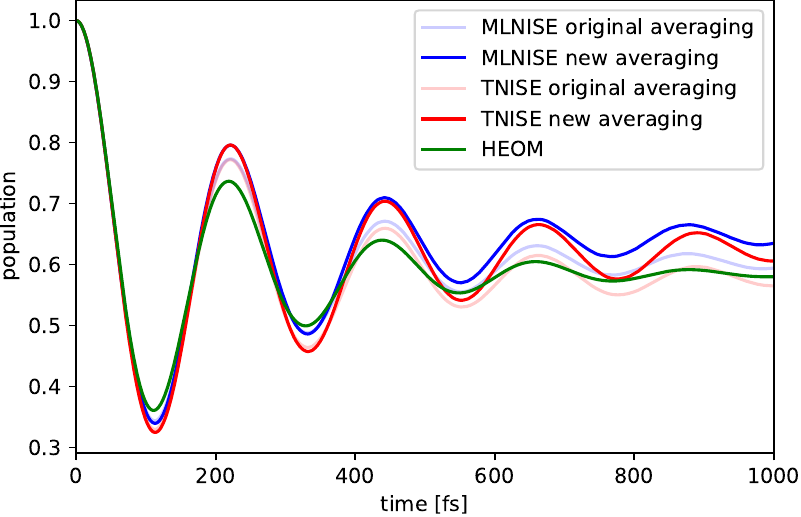}
}
\\[-4ex]
\subfloat[][]{
\includegraphics[width=0.485\textwidth]{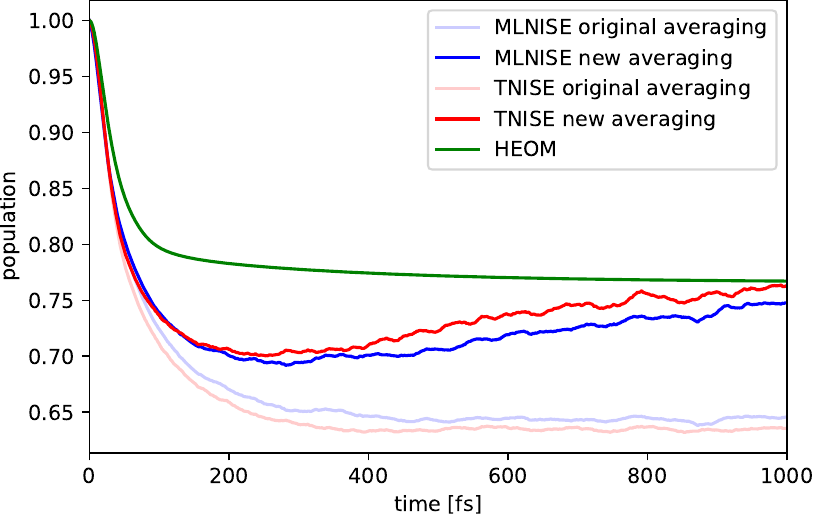}
}
\hskip0em\relax
\subfloat[][]{
\includegraphics[width=0.485\textwidth]{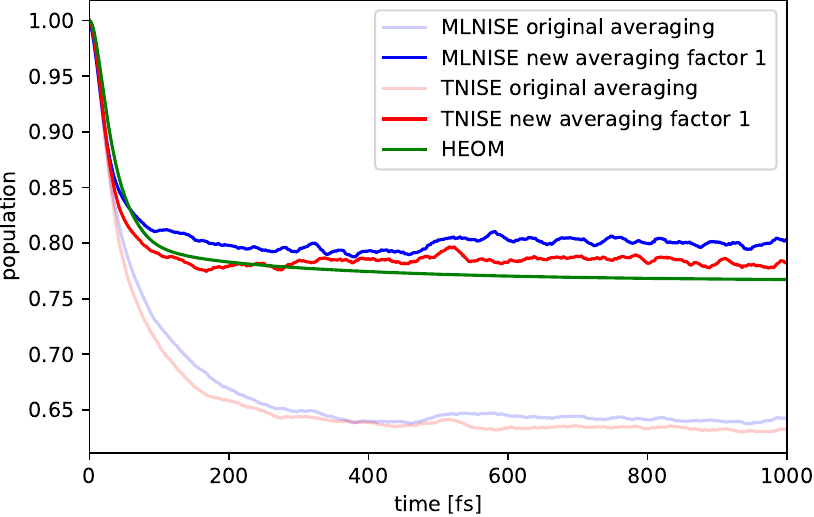}
}
\caption{Example population dynamics of two-site systems with the NISE trajectories averaged over 10,000 realizations. The parameters can be found in Tab.~\ref{table:sampleSytemParameters}. Panel a) shows a case where the new averaging outperforms the original averaging, while in panel b) the opposite is true. In panel c)  an example is delineated in which the lifetime with the empirical factor of five from Eq.~\ref{eq:empirical_factor} is too high, while panel d) shows the same system with the factor set to one.}
\label{fig:new_averaging}
\end{figure}

For the rest of the section, we will focus on the overall performance across a diverse range of 2-site systems, which is critical to demonstrate the robustness of the method and to avoid biases from cherry-picked examples. Using the methodology described in Ref.~\onlinecite{holt23a}, we analyzed the efficacy of both the TNISE and the MLNISE schemes with the new and the original averaging schemes across 10,000 sample systems. The performance was evaluated as the MSE for a 1~ps-long population dynamics simulation, compared to HEOM results obtained using the PyHeom package \cite{iked20a}. The neural network from Ref.~\onlinecite{holt23a} that was trained on 2-site systems was employed for the MLNISE approach and was not trained specifically using the modified averaging procedure. Each system utilizes a Drude spectral density with parameters randomly selected from the range specified in Tab.~\ref{table:sampleSytemParameters}. The performance is depicted in Fig.~\ref{fig:new_averaging_many_systems}, showing the MSE as a function of the reorganization energy $\lambda$ of the respective systems. The graph illustrates that except for low values of $\lambda$  where the new averaging underperforms, it significantly surpasses the original averaging, especially as $\lambda$ increases. The underperformance for low $\lambda$ is not that surprising, as these systems do not reach thermal equilibrium within the 1~ps time scale of the investigated trajectories, as can be seen, for example,  in Fig.~\ref{fig:new_averaging}b. Hence, the discrepancies at short times outweigh the improved thermal equilibrium in these cases.

Figure~\ref{fig:new_averaging} offers a deeper insight into specific cases, reinforcing the general findings. In panel a, the new averaging method performs exceptionally well, closely mirroring the reference HEOM results, thus more accurately reflecting the thermal equilibrium. Panel b demonstrates an instance of underperformance in the low reorganization energy regime, where the transition to the new averaging starts too early. The system clearly has not reached equilibrium, hence the long-term limit required by Eq.~\ref{eq:averagingissue} is not achieved while the system already transitions to the modified averaging scheme. In contrast, panel c displays a scenario where a faster switching to the new averaging would have been beneficial, as the system has already reached equilibrium at an earlier time. Panel d depicts the results for a modified empirical factor in Eq.~\ref{eq:empirical_factor}, i.e., a value of 1 instead of 5, illustrating that an earlier transition to the new averaging improves outcomes in such cases.
These case studies underline the importance of accurate lifetime estimations in the new averaging method. While the empirical factor of 5 generally benefits the overall performance across the 10,000 test system, it can be overly conservative or excessive in specific instances, such as those depicted in Fig.~\ref{fig:new_averaging}b, c, and d.

Overall, the comprehensive analysis with extensive simulations confirm that the modified averaging procedure substantially improves the population dynamics, especially when reaching the thermal distribution in the long-time limit. Thus, these findings validate the intuition derived from Eq.~\ref{eq:averagingissue} and show a marked improvement over the original ensemble averaging. Further improvements can be envisioned by refining the method for the lifetime predictions and by replacing the simple exponential fit. Moreover, a newly trained neural network for the MLNISE approach trained specifically for the modified averaging procedure would certainly be beneficial.
\begin{table}[!bt]
\begin{small}
\begin{center}
 \begin{tabular}{c c c c c c} 
 \hline
 \hline
  System &  $V$[cm$^{-1}$] & $\tau$[fs] & $\lambda$[cm$^{-1}$] &$E$[cm$^{-1}$] & $T$[K] \\ [0.5ex] 
 \hline\hline
 Fig.~\ref{fig:new_averaging_many_systems} & [-200, \ldots, 200] & [10, \ldots, 200] & [0, \ldots, 500] & [-500, \ldots, 500] & 300 \\
  \hline
 Fig.~\ref{fig:new_averaging}a & 134.8 & 81.8 & 192.6 &  -302.4 & 300 \\
   \hline
 Fig.~\ref{fig:new_averaging}b & 63.5 & 81.0 & 9.2 &  -72.3 & 300 \\
   \hline
 Fig.~\ref{fig:new_averaging}c\&d & -82.4 & 65.8 & 386.3 &  -253.5 & 300 \\
 \hline
 \hline\hline
\end{tabular}
\end{center}
\end{small}
 \caption{Parameter ranges used for the benchmark in Fig.~\ref{fig:new_averaging_many_systems} and parameters for the sample systems in Fig.~\ref{fig:new_averaging}. For Fig.~\ref{fig:new_averaging_many_systems}, the energy range specifies the range from which the energies $E_1$ and $E_2$ are chosen, while for the other figures $E_2$=0 is set to zero and the listed energy specifies  $E_1$. The parameter values for the sample two-site systems were rounded for readability.}
 \label{table:sampleSytemParameters}
\end{table}

\section{Generation of Structured Noise}
\label{sec:NoiseGen}
\subsection{Introducing the Algorithm}

The generation of random numbers with a computer today is usually still based on a
pseudo-random number generator, while  it has become possible to use special
hardware to produce quantum random numbers that cannot be predicted
\cite{mann23a}. The differences between quantum and pseudo-random numbers are,
however, not the topic of the current study, assuming the use of a “good” random
number generator. Here, the objective is to generate time series of the site energy
fluctuations obeying a pre-defined behavior of the autocorrelation function, i.e., also spectral density (see Eq.~\ref{eq:jw}). 
In the preceding section, we employed noise with exponentially decaying autocorrelation functions, resulting in Drude spectral densities. This functional form of the autocorrelation function was chosen since the corresponding noise can be generated rather easily, as, for example, described in Ref.~\onlinecite{fox88a} and since it has been used in previous quantum dynamics studies \cite{jans18a,aght12a,holt23a}. 
For more comprehensive investigations, it is, however,  very valuable to be able to generate structured noise that adheres to more general autocorrelation functions and corresponding spectral densities. This versatility enables the application of noise generation techniques to a wider range of scenarios, including those where experimental spectral densities should be mimicked.

The creation of white noise which is delta-correlated in time and
uniformly distributed in frequency is a standard operation.  
Producing a series of random numbers which is not
delta-correlated in the time domain is, however, a bit more involved. Procedures to generate series of noise with more complicated characteristics have been used in 
several fields already including, for example, computer graphics \cite{mast87a} and computational geography \cite{rava00a}. 
In the area of molecular simulations, approaches to generate correlated noise have also been reported \cite{dijk09a, hart16a}.  
In the context  of the Hierarchy of Pure States (HOPS) approach \cite{hart16a,hart17a}, two methods to
generate complex valued random numbers with arbitrary time autocorrelation
functions have been proposed.  One of the two approaches is based on the
Karhunen-Loève expansion \cite{font12a}, while the other one employs the Fast Fourier
Transfomation (FFT) \cite{cool65a,bloo00a} method. 
In the present study, a procedure for generating a
time series of real valued random numbers is needed which can produce a
specific but potentially very complicated time autocorrelation functions. Actually, many of the mentioned algorithms utilize the FFT and are very similar to the “circulant embedding of the covariance matrix” algorithm described in Ref.~\onlinecite{diet97a}, although they often differ in small but significant ways. Because the same holds true for our approach, we briefly show details of our variant to be clear how we generated the correlated noise to improve reproducibility.

The aim is to generate structured real noise $\eta^T(t)$ which we term the target noise. This noise should correspond to a specific spectral density, i.e., to a specific Fourier transform of the noise correlation function according to  Eq.~\ref{eq:jw}.  
White noise $\eta^W(t)$ is characterized by  a delta-correlated (normalized) autocorrelation function $C^W(t)=\langle \eta^W(t)  \eta^W(0) \rangle =\delta(t)$, where an average over sufficiently  many noise realizations has to be performed, and its Fourier transform equals unity, i.e,  $\tilde{C}^W(\omega)=1$. 
The target noise  $\eta^T(t)$  has the time correlation function   
\begin{equation}
\label{eq:autocorrelation_continous}
	C^T(t)=\int\limits_{-\infty}^\infty \langle \eta^T(\tau) \eta^T(t+\tau) \rangle d\tau~.
\end{equation}
According to the Wiener-Khinchin theorem its Fourier transform is simply
\begin{equation}
 \tilde{C}^T(\omega) =\langle|\tilde{\eta}^T(\omega)|^2  \rangle=\langle|\tilde{\eta}^T(\omega)|^2  |\tilde{\eta}^W(\omega)|^2  \rangle
 =\langle |\tilde{\eta}^T(\omega) \tilde{\eta}^W(\omega)|^2  \rangle~.  \label{ctft}
\end{equation}
Since $\tilde{C}^T(\omega)$ is non-negative, this equation is fulfilled by 
\begin{equation}
\tilde{\eta}^T(\omega) =  \sqrt{\tilde{C}^T(\omega)}  \tilde{\eta}^W(\omega) \label{etatft}
\end{equation}
or in the time domain using the Fourier back transform  $\mathcal{F}^{-1}$
\begin{equation}
	\eta^T(t) =  \mathcal{F}^{-1}\left[ \sqrt{\tilde{C}^T(\omega)}  \tilde{\eta}^W(\omega) \right](t)~.   \label{etat}
\end{equation}
In terms of an algorithm, one first draws a series of random numbers from a Gaussian distribution (or whatever distribution is desired in energy space) and based on a time step $\Delta t$ makes it an equidistant time series $\eta^W(t)$. This time series is then Fourier transformed and element-wise multiplied by $\sqrt{\tilde{C}^T(\omega)}$ which characterizes the target noise. Subsequently, a Fourier back-transformation is performed to obtain the target noise $\eta^T(t)$. 

\subsection{Applicability of the Algorithm}

\begin{figure}[tb]
\centering
\includegraphics[clip,width=13cm]{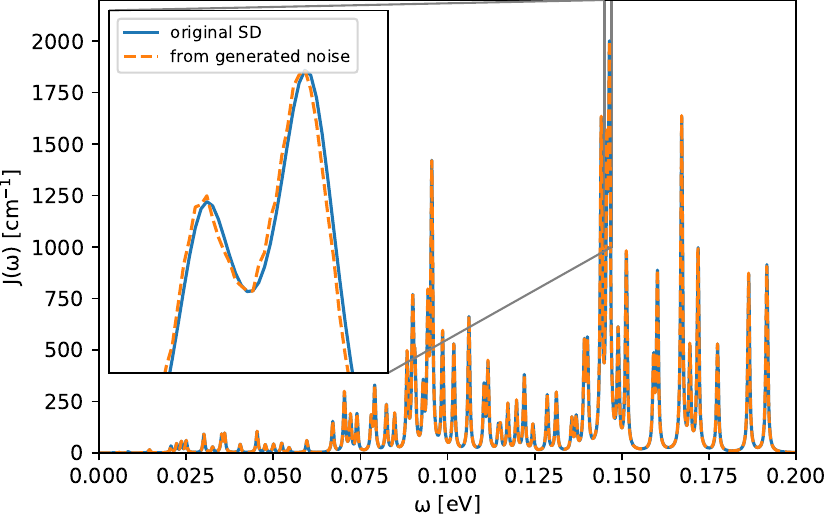}
\caption{Reconstruction of the experimental spectral density of the FMO complex \cite{raet07a,mait20a} by using a 100~ps-long trajectory with a 10~fs time step and 100,000 realizations.}
\label{fig:realizations_fullSD}
\end{figure}
\begin{figure}[!ht]
\captionsetup[subfloat]{position=top,singlelinecheck=false,justification=raggedright,labelformat=brace,font=bf}
   \centering
   \subfloat[][]{
   \includegraphics[width=0.485\textwidth]{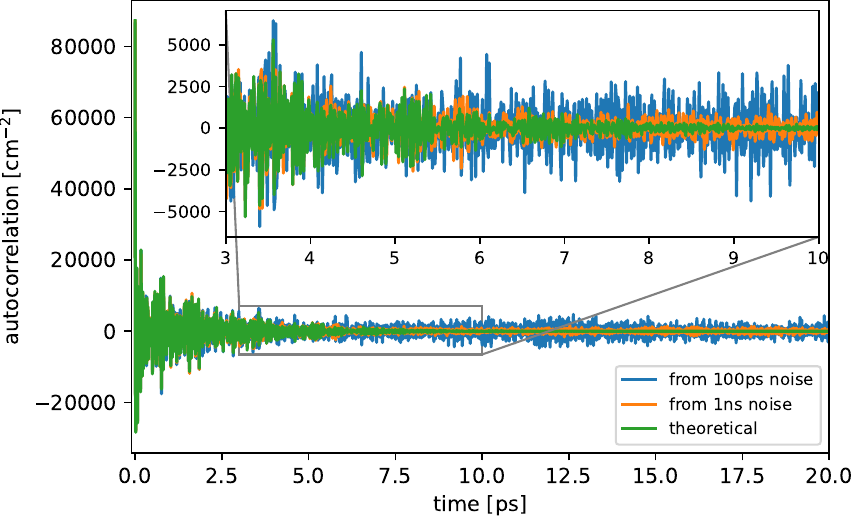}
    }
   \hskip0em\relax
   \subfloat[][]{
   \includegraphics[width=0.485\textwidth]{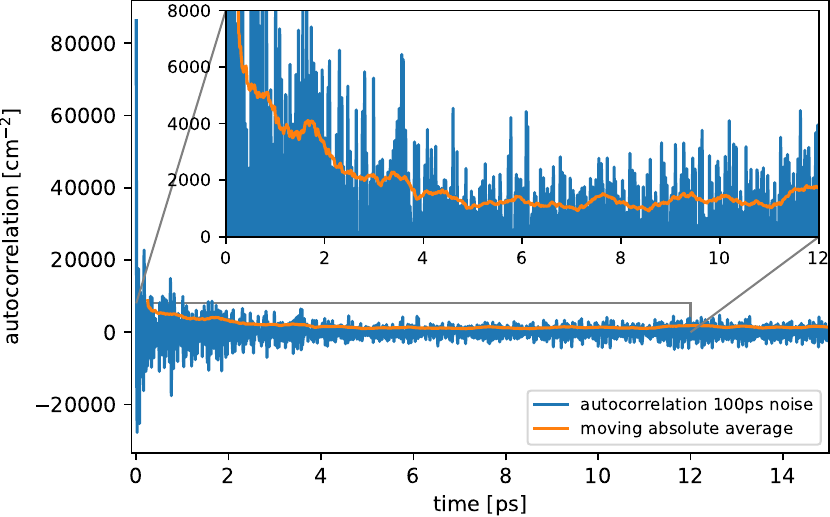}
    }
  \caption{Autocorrelation function of the experimental spectral density of the FMO complex (see Fig.~\ref{fig:realizations_fullSD}). The left panel shows the theoretical autocorrelation function calculated by inverting Eq.~\ref{eq:jw} compared to the autocorrelation functions obtained from Eq.~\ref{eq:autocorrelation_from_noise} for a 1~ns and a 100~ps-long  trajectory. The right panel again shows the autocorrelation obtained from the 100~ps-long trajectory, but in this panel together with a 500 fs moving average over the absolute value of the same autocorrelation function.}
  \label{fig:autocorrelation}
\end{figure}

Demonstrating the efficacy of the noise generating algorithm can be achieved by reproducing the spectral density from the generated noise. To do this, a highly structured experimental spectral density was reproduced with many realizations of long trajectories of energy fluctuations (noise). This step validates that trajectories generated by the algorithm follow all relevant spectral features for reasonable time steps and trajectory lengths.
The spectral density $J(\omega)$ and the autocorrelation $C(t)$ of the generated noise are inherently linked via a cosine transform (see Eq.~\ref{eq:jw}).
Detailed computational considerations for calculating this cosine transform are outlined in the Appendix \ref{app:Cosine_Transform}.
The bath autocorrelation function in the discrete domain $C(t_j)$ can be calculated from the noise trajectory $E(t_i)$ in a discrete version of Eq.~\ref{eq:autocorrelation_continous} as \cite{damj02a}
\begin{equation}\label{eq:autocorrelation_from_noise}
    C(t_j) = \frac{1}{N-j} \sum_{i=1}^{N-j} \Delta E(t_i + t_j) \Delta E(t_i)~.
\end{equation}
Here $N$ denotes the number of time steps in the trajectory. For longer correlation times, the sum contains less and less terms, hence it is common practice to only trust the values with $j<N/2$. Since this autocorrelation is calculated for a finite window, the autocorrelation does not decay completely to zero as expected for longer times, as can be seen in Fig.~\ref{fig:autocorrelation}. This shortcoming can be mitigated by sampling longer trajectories or multiple realizations. However, contrary to what is sometimes done in other studies, 
subdividing the individual trajectory into overlapping smaller windows does not reduce the respective error. 
The autocorrelation at time $t_j$ is already averaged over all possible value pairs $(\Delta E(t_i + t_j), \Delta E(t_i))$ along the trajectory. Hence, subdividing the trajectory (and also using overlapping windows \cite{loco18b,mait23a}) does not add further energy pairs that would improve the average at $C(t_j)$.
Another approach for solving the problem of the non-decaying autocorrelation is to apply a damping function \cite{loco18b}. Details of such an approach will be discussed in the next section.

In a first step, we only aim to verify the applicability of the noise-generating algorithm for a realistic spectral density. To do so, the problem of the non-decaying autocorrelation function is solved by performing an average over many realizations. This approach is possible since the algorithm allows for a computationally cheap generation of many independent realizations.
Fig.~\ref{fig:realizations_fullSD} shows an almost perfect match of the original spectral density and the one calculated from a noise trajectory. Thus, for reasonable time steps and trajectory lengths, the above detailed algorithm is able to generate structured noise capturing all features of a highly structured spectral density. To illustrate the applicability for systems of interest, the original spectral density was chosen to be an experimentally determined spectral density of the Fenna–Matthews–Olson (FMO) complex \cite{raet07a,mait20a}. 

\section{\label{sec:SD-reconstruction}Calculating Spectral Densities using a  Limited Amount of Data}

The algorithm for generating structured noise provides the ability to generate almost unlimited noise. If the site energy fluctuations (noise) are based, however, on numerically much more demanding calculations such as excited state calculations along MD or QM/MM trajectories, the 
noise generation is computationally expensive, resulting in a limited amount of energy values being available \cite{cign22a,mait23a}. Consequently, the present study also focuses on establishing best practices and assessing the accuracy of spectral densities obtained from noise trajectories of limited length. This assessment is done by utilizing the known correct spectral density as a reference for the generated limited noise. Since the spectral density is basically a one-sided Fourier transform of the autocorrelation function as given by Eq.~\ref{eq:jw}, the Nyquist-Shannon sampling theorem \cite{nyqu28a,shan49a,unse00a} plays a major role when analyzing the frequency range of the spectral density. According to this theorem, 
the largest angular frequency one can accurately resolve is given by  
\begin{equation}
    \omega_{\text{nyq}}=2 \pi\frac{f_{\text{s}}}{2}=\frac{\pi}{\Delta t} ,
\end{equation}
where $f_{\text{s}}$ denotes the sampling frequency and $\Delta t$  the time step. From a numerical point of view, it makes sense to choose the time step as large as possible while still resolving all relevant frequencies.
With the application of light-harvesting complexes in mind, a time step $\Delta t$  of 10~fs is sufficient, as it can resolve the internal frequencies of chlorophyll molecules, which reach up to about 0.2~eV.
As discussed in the previous section, the biggest issue in calculating the autocorrelation function for limited times is that it does not decay to zero. For cases in which the energy fluctuations are based on excited state calculations, it is often not feasible to calculate very long trajectories or a larger number of independent realizations. In such cases, it makes sense to use a damping function for the autocorrelation function. Previously, Gaussian $\exp(-(t/t_c)^2)$ or exponential $\exp(-t/t_c)$ damping functions have been suggested \cite{olbr10a,loco18b} to force the autocorrelation down to zero a times larger than $t_c$ as they  result in Gaussian or Lorentzian broadening of the spectral density, respectively.
Another option is to cut the autocorrelation function when it is close to zero and pad it with zeros, since the frequency resolution is inversely proportional to the length of the autocorrelation function \cite{mait20a,mait21a}. 
This procedure is equivalent to using a step function with $t_c$ as the cutoff time point for the damping function.
As an aside, one could consider more general damping functions of the form $\exp(-(t/t_c)^b)$.
All three methods are special cases of this general damping function with the step function corresponding to the limit of $b$ going to infinity.
Rather than padding the autocorrelation function with zeros, it is also possible to increase the frequency resolution with super-resolution techniques \cite{wei14a}, where, instead of using an FFT, the spectral density is determined by fitting the autocorrelation function to terms corresponding to known spectral density terms. One such technique has, e. g., been used to fit the autocorrelation function to a number of Drude-Lorentz peaks \cite{mark16a}. We tested this method and found that it performed very poorly with our reference spectral density without modification. With a few alterations, we could achieve reasonable results. However, we found that the additional effort was not worth it compared to the much simpler FFT-based methods and decided to exclude these results from the main section. The results of the super-resolution technique as well as the details of the modifications can be found  in Appendix~\ref{app:super_resolution}.

\begin{figure}
\captionsetup[subfloat]{position=top,singlelinecheck=false,justification=raggedright,labelformat=brace,font=bf}
   \centering
   \subfloat[][]{
   \includegraphics[width=0.485\textwidth]{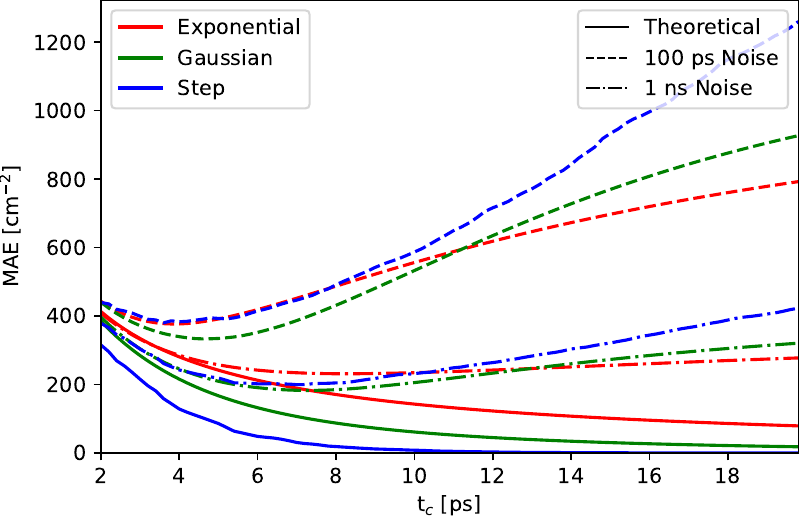}
    }
   \hskip0em\relax
   \subfloat[][]{
   \includegraphics[width=0.485\textwidth]{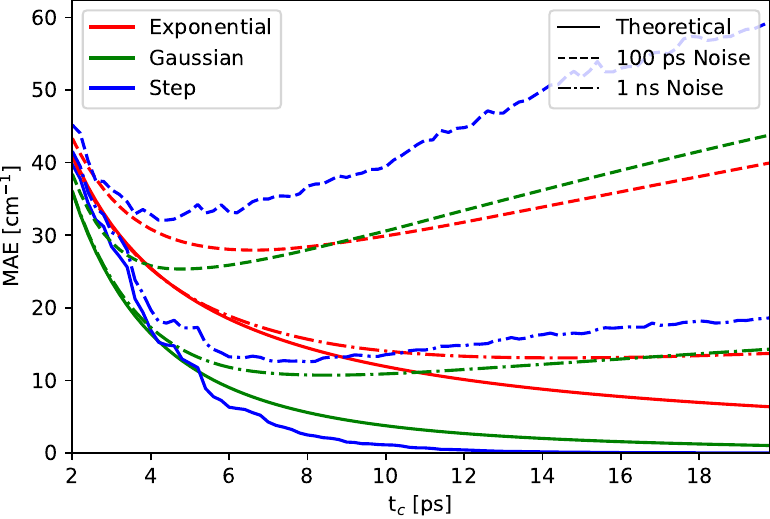}
    }
  \caption{Mean absolute error analysis of different damping methods relative to the theoretical spectral density as a function of the cutoff time $t_c$. The left panel shows the MAE derived from the initial 20~ps of the autocorrelation function, while the right panel addresses the MAE associated with the spectral density. The different colors correspond to three damping strategies: utilizing a step function, exponential damping, and Gaussian damping. The continuous lines represent errors calculated from the theoretical autocorrelation function, the dashed lines from autocorrelation derived from a 100~ps-long noise trajectory, and the dash-dotted lines from autocorrelation obtained from a 1~ns-long noise trajectory.}
  \label{fig:errors}
\end{figure}

In the following, the three FFT-based methods, Gaussian and exponential damping as well as the step function will be investigated in more detail by assessing their performance with regard to the same spectral density as in the previous section. Fig.~\ref{fig:errors} delineates the Mean Absolute Error (MAE) introduced by the application of the three approaches to dampen the autocorrelation function across different cutoff times. The MAE is assessed under three distinct scenarios: when applied to the theoretical autocorrelation function, derived by inverting Eq.~\ref{eq:jw} (see also Appendix \ref{app:Cosine_Transform}), and when applied to autocorrelation functions obtained from noise trajectories of 100 ps and 1 ns durations. Specifically, Fig.~\ref{fig:errors}a focuses on the MAE compared to the theoretical autocorrelation functions. Since the autocorrelation functions decay quickly as shown in Fig.~\ref{fig:autocorrelation}, the comparison is only determined for the first 20 ps. Hence, in the theoretical case, the step function reaches a MAE of zero for $t_c = 20 \text{ps}$. In general, the step function has the lowest MAE for all values of  $t_c$ in the theoretical case, followed by the Gaussian damping and the exponential damping having the highest MAE. For the auto-correlation functions obtained from the noise trajectories, the Gaussian damping achieves the lowest MAE in both cases. In the 1~ns case the step function achieves a very similar result with the exponential dampening result being slightly worse. Both the step function and the exponential attenuation are slightly worse in the 100~ps case. In addition, Fig.~\ref{fig:errors}b examines the MAE pertaining to the spectral density calculated from the damped autocorrelation functions. The analysis reveals that for the theoretical autocorrelation function, the Gaussian and step-function schemes reach an accuracy superior to that  of the exponential method. However, in the context of autocorrelation functions derived from actual noise trajectories, the Gaussian approach reaches the lowest MAE, affirming its efficacy in managing noisy data.

\begin{figure}
\captionsetup[subfloat]{position=top,singlelinecheck=false,justification=raggedright,labelformat=brace,font=bf}
   \centering
   \subfloat[][]{
   \includegraphics[width=0.485\textwidth]{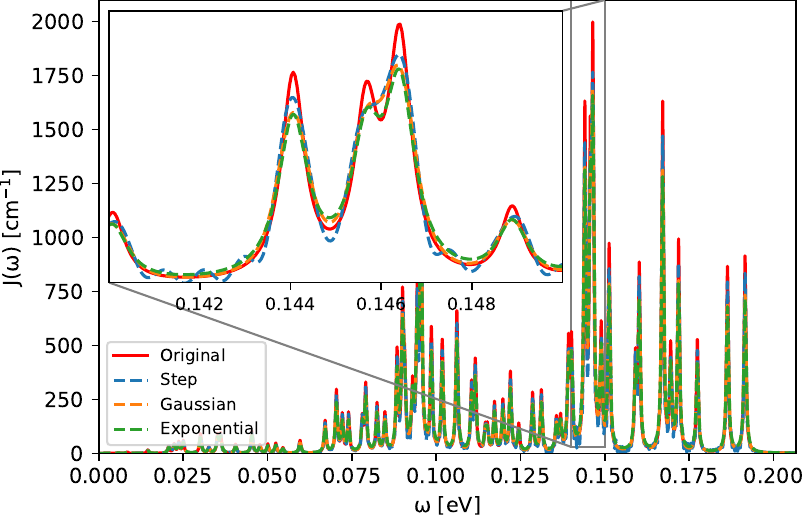}
    }
   \hskip0em\relax
   \subfloat[][]{
   \includegraphics[width=0.485\textwidth]{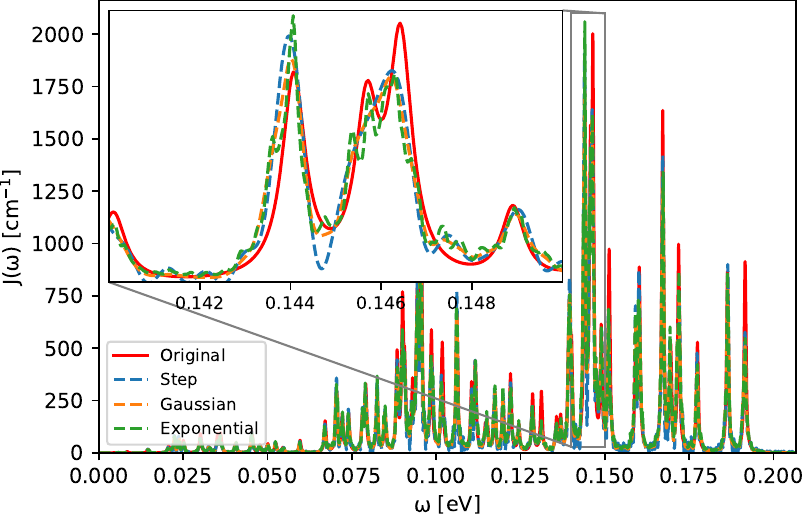}
    }
  \caption{Comparison of spectral densities derived from autocorrelation functions damped in different ways with the benchmark spectral density. Cutoff times of 5 ps for the Gaussian damping and the step function were chosen while for the exponential damping this values is set to 7.5 ps based on the optimal performance displayed in Fig.~\ref{fig:errors}. Panel a) illustrates spectral densities obtained from theoretical autocorrelation functions, panel b) showcases spectral densities derived from noisy autocorrelation functions of a 100~ps-long trajectory.}
  \label{fig:damping_Spectral_density}
\end{figure}

A further investigation is presented in Fig.~\ref{fig:damping_Spectral_density}a, which contrasts the various spectral densities derived from autocorrelation functions subjected to different damping schemes compared to the original FMO spectral density. The cutoff times  were chosen to be 5~ps for Gaussian damping and step function as well as 7.5~ps for the exponential damping. These times were chosen based on their performance for the autocorrelation functions based on the 100 ps noise, as indicated in Fig.~\ref{fig:errors}b. The comparative analysis elucidates that the spectral density using Gaussian damping aligns closest with the overall spectral density. At the same time, the exponential damping function resolves individual peaks slightly better, at the cost of a worse overall alignment. The step-function approach, while capturing the peak heights more accurately, introduces nonphysical oscillations as artifacts.
An analogous comparison for noise-based autocorrelation function obtained from a 100 ps time series is showcased in Fig.~\ref{fig:damping_Spectral_density}b. These findings mirror those based on the “exact” autocorrelation function, with Gaussian damping maintaining the closest alignment to the original spectral density. While exponential damping still exhibits a refined peak resolution, it also mistakenly accentuates noise artifacts as additional peaks. This aspect becomes evident in the misinterpretation of the double peak at 0.146 eV as a triple peak. On the contrary, the Gaussian damping and step-function approaches resolve the double peak as a single peak with a slight shoulder.
In conclusion, a Gaussian damping function emerges as the most accurate strategy overall for processing spectral densities derived from limited noise data. The exponential damping and the step-function schemes might still be useful for specific tasks: A better resolution of  peaks in case of the exponential damping and an improved matching of peak heights for the step-function scheme. 

Above, we have determined the cutoff time based on the performance with respect to the theoretical spectral density. Considering most applications where the noise is based on, e.g., MD simulations, such a theoretical spectral density is unknown. In the following, a slightly more practical approach is tested by determining the cutoff time as the time
at which the autocorrelation function ceases to exhibit an exponential decay and begins to be predominantly influenced by unsystematic noise. This approach is substantiated by an observation in Fig.~\ref{fig:errors} that the results are not very sensitive to the cutoff time, i.e., cutoff times  slightly above or below the optimal value still yield satisfactory results. As can be seen in Fig.~\ref{fig:autocorrelation}b,  employing a 500 fs running average over the absolute value of the autocorrelation function simplifies 
the identification of the crossover region between exponential decay of the autocorrelation function and 
a noise-dominated flat regime. For the present example, this crossover region nicely coincides with the previously identified optimal cutoff time of roughly 5~ps, thereby giving some credibility to the proposed procedure.










\section{\label{sec:population_dynamics}Population Dynamics with Highly Structured Spectral Densities}

In the previous sections, a modified ensemble-averaging procedure for the TNISE approach was proposed and a scheme for generating arbitrary structured noise was detailed and tested. A primary objective of this paper, particularly through these two sections, is to facilitate NISE and TNISE simulations with arbitrary, potentially experimental spectral densities. Previous studies employed either simplified spectral densities \cite{aght12a,jans18a,holt23a} or noise trajectories derived directly from MD or QM/MM simulations \cite{mait20a,sarn24a}. With the methods presented in this work, it is possible to contrast population dynamics from MD noise trajectories and generated noise derived from experimental spectral densities. Such a comparison might provide additional insights into problems described by quantum dynamics, e.g., exciton transfer pathways.

Before delving into examples of such population dynamics, we will shortly point out two potential issues connected to the population dynamics for systems with structured noise.
Firstly, a time step of 10~fs  was suggested for systems where the spectral density is of interest up to energies of 0.2~eV, such as molecules involving double bonds. Such a large time step is problematic for the population dynamics of such systems. 
If higher frequencies are absent in the spectral density (which was the reason for the large time step in the first place), this issue can be addressed by resampling the time series to a smaller time step (see Appendix~\ref{app:timeseries_resampling}). Secondly, while overlapping windows were deemed ineffective for spectral density calculations in the discussions above, they are indeed useful for population dynamics. 

\begin{figure}[!htb]
\captionsetup[subfloat]{position=top,singlelinecheck=false,justification=raggedright,labelformat=brace,font=bf}
\subfloat[][]{
\includegraphics[width=0.485\textwidth]{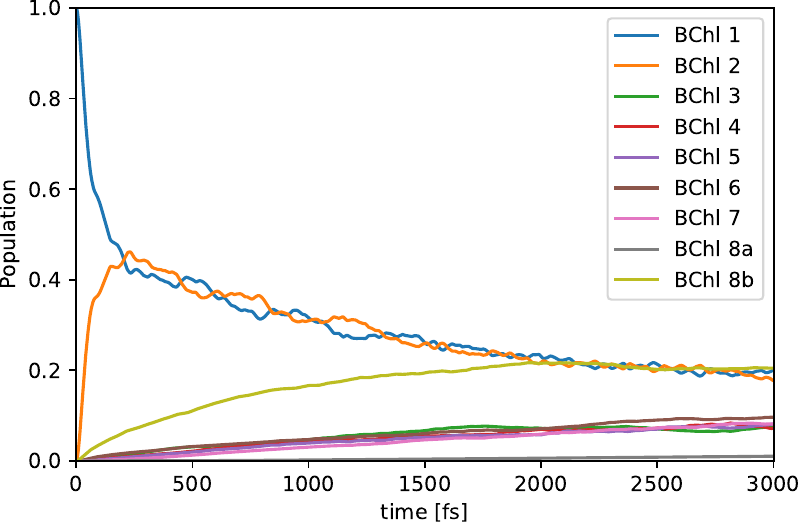}
}
\hskip0em\relax
\subfloat[][]{
\includegraphics[width=0.485\textwidth]{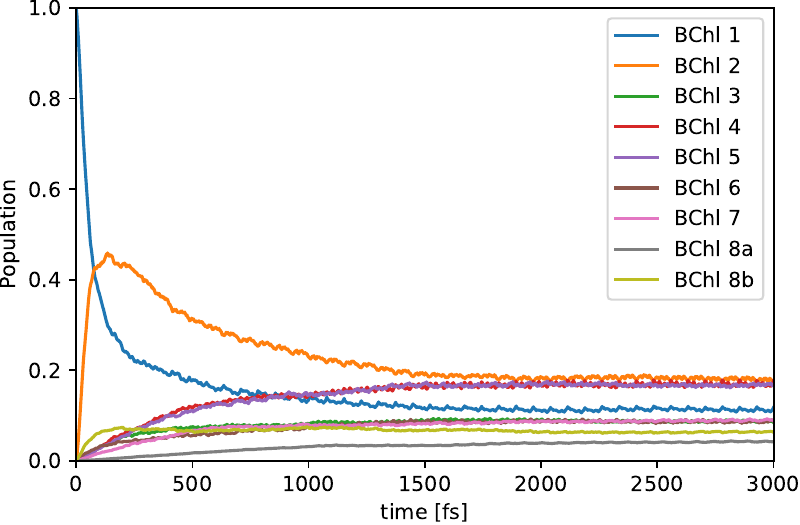}
}
\\[-4ex]
\subfloat[][]{
\includegraphics[width=0.485\textwidth]{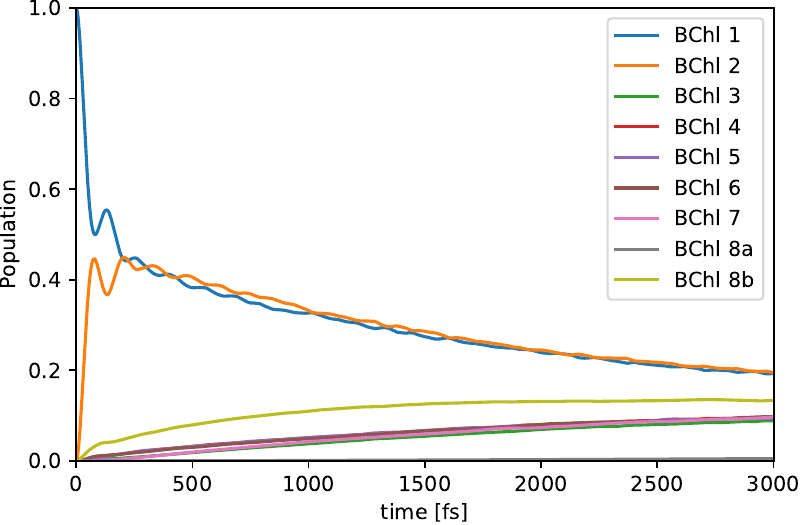}
}
\hskip0em\relax
\subfloat[][]{
\includegraphics[width=0.485\textwidth]{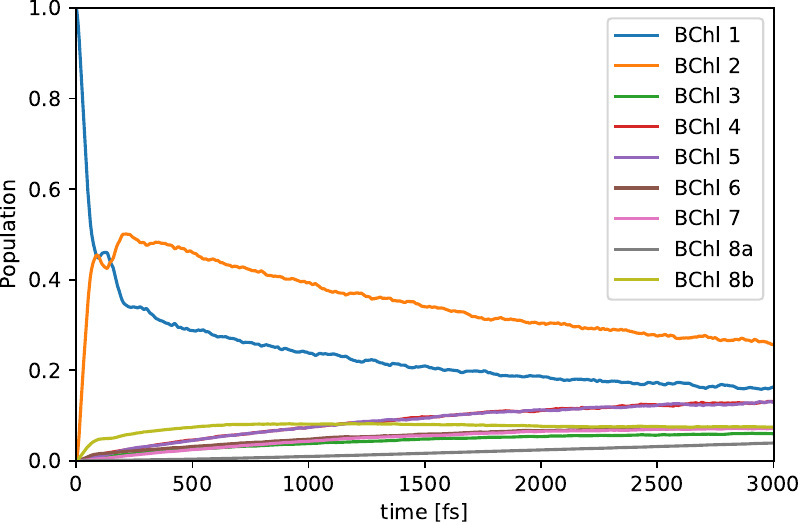}
}
\caption{Population dynamics of FMO. Panel a shows the results of the NISE approach and panel b the TNISE results utilizing QM/MM noise trajectories from Ref.~\onlinecite{mait20a}, while panel c displays the NISE and d the TNISE results utilizing the noise generation algorithm with the experimental spectral density as shown in  Fig.~\ref{fig:realizations_fullSD}. }
\label{fig:population_FMO}
\end{figure}

Returning to the main topic of population dynamics, the FMO complex continues to serve as our system of interest. As an example use case, we contrast population dynamics derived from QM/MM noise trajectories taken from \cite{mait20a}, with those based on generated noise from experimental spectral densities, specifically the same spectral densities as used above and shown in Fig.~\ref{fig:realizations_fullSD}. To ensure comparable conditions, the time-averaged Hamiltonian from the QM/MM noise trajectory is employed as the system Hamiltonian for the generated noise. The off-diagonal coupling elements remain constant for both cases, as detailed in Ref.~\onlinecite{mait20a}.
Since machine learning models in general are limited by their training range, the MLNISE models trained in Ref.~\onlinecite{holt23a} are not suitable for this example. These models were trained on simpler noise profiles following Drude spectral densities and for dynamics not exceeding 1 ps. Training models for more complicated noise and longer trajectories, while interesting, is outside the scope of this paper. 
Instead, only the NISE approach and its thermalized variant, TNISE, are applied to the FMO complex to study exciton dynamics over a 3 ps time interval after initially exciting the  bacteriochlorophyll 1 (BChl 1). The modified averaging scheme introduced in Section~\ref{sec:new_averaging_procedure} has been used for the TNISE approach. Figure~\ref{fig:population_FMO} illustrates the population dynamics for different approaches: Panels a and b show the population dynamics based on the 40 ps QM/MM noise trajectory for NISE and TNISE schemes, respectively. The trajectory is split into 500 overlapping realizations with a 74 fs stride between windows. Panels c and d present the dynamics using generated noise following the experimental spectral density, employing the NISE and TNISE approaches, respectively. A considerable advantage for noise generated based on experimental spectral densities is that the computational cost of more realizations is minimal compared to the costly generation of longer QM/MM trajectories. Hence, 5,000 realizations were utilized for this approach, resulting in less noisy dynamics.
The different methods show partially significant differences in the population dynamics. One example is the population transfer towards BChl 8b, noted in Ref.~\onlinecite{mait20a}, which is markedly reduced when using experimental spectral densities and the TNISE scheme as can be seen in Fig.~\ref{fig:population_FMO}c and d. Interpreting such qualitatively different results can provide new insights, highlighting the utility of the introduced methods. 
Regarding thermalization, as can be seen in Fig.~\ref{fig:population_FMO}b, the TNISE scheme based on the 40 ps QM/MM trajectory reaches a steady state before 2 ps, a surprisingly early convergence that is not observed for the (non-thermalized) NISE scheme within the same timeframe. This result prompts a further investigation into whether this rapid thermalization is an artifact of the TNISE procedure. We note that this rapid thermalization occurs regardless of the new or old averaging procedures being used. The lifetimes are estimated based on the (non-thermalized) NISE scheme, which reaches the  its long time limit at much later times. Significant interpolation towards the new averaging scheme happens later than the shown timescale. (See also Fig.~\ref{fig:new_averaging_FMO})

\begin{figure}
\captionsetup[subfloat]{position=top,singlelinecheck=false,justification=raggedright,labelformat=brace,font=bf}
   \centering
   \subfloat[][]{
   \includegraphics[width=0.485\textwidth]{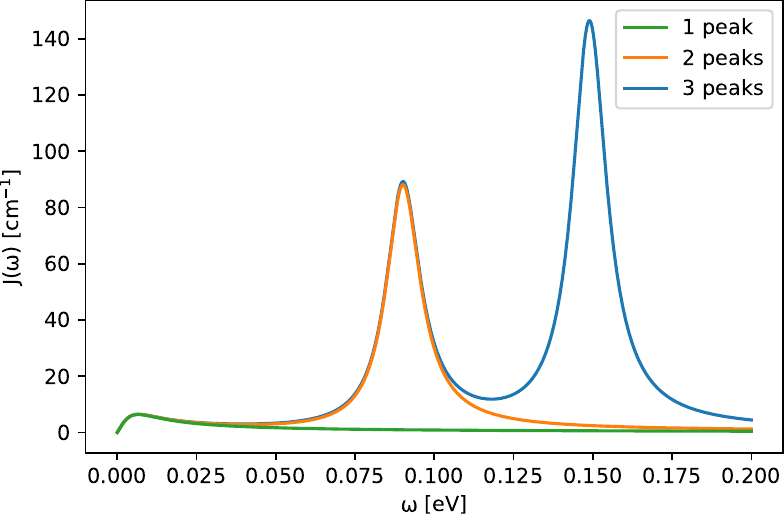}
    }
   \hskip0em\relax
   \subfloat[][]{
   \includegraphics[width=0.485\textwidth]{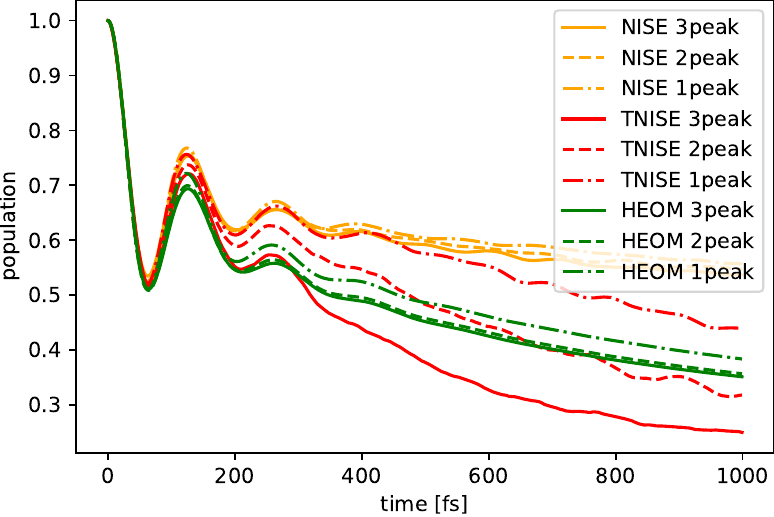}
    }
  \caption{Illustration of the effect of high frequencies in the spectral density on the population dynamics. Panel a) shows an artificial spectral density with three peaks, while panel b) displays the respective population dynamics for the NISE, TNISE, and HEOM approaches. 10,000 realizations were used for the NISE variants.}
  \label{fig:high_freq_thermalization_issue}
\end{figure}

For this purpose, we conducted an analysis alongside the numerically exact HEOM formalism. Due to  computational demands of the HEOM scheme, which scales unfavorably with both the number of bath modes and system size, we designed a simpler spectral density with three broad peaks retaining a similar frequency range compared to the experimental spectral density as shown in  Fig.~\ref{fig:high_freq_thermalization_issue}a.  The form of the simpler spectral density is adapted from the GPU-HEOM tool \cite{krei13b}, which was used for the HEOM calculations in this section 
\begin{equation}
J(\omega)= \frac{1}{\pi} \sum_{k=1}^n \left [  \frac{\nu_k\lambda_k\omega}{\nu_k^2 + (\omega-\Omega_k)^2}+ \frac{\nu_k\lambda_k\omega}{\nu_k^2 + (\omega+\Omega_k)^2} \right ] ~ .
\end{equation}
The spectral peaks are positioned at $\Omega_k = 0\text{,} ~ 725 ~ \text{and} ~ 1200 ~ \text{cm}^{-1}$, each having a reorganization energy of  $\lambda_k = 20 ~ \text{cm}^{-1}$ and a peak width of $\nu_k = 100 ~ \text{fs}$.
For computational feasibility and easier comparison, a simple two-site Hamiltonian with an energy difference of $\Delta E = 200 ~ \text{cm}^{-1}$ and a coupling of $V = 100 ~ \text{cm}^{-1}$ was chosen.
The population dynamics for the different mentioned approaches is depicted in 
Fig.~\ref{fig:realizations_fullSD}b) for spectral densities with one, two, and three peaks. 
For the HEOM formalism, but also the original NISE approach, the highest frequency peak in the spectral density has only a very small effect on the present population dynamics, which is not surprising since the frequency of these bath modes is quite far away from the transition energies in the system.  For the TNISE scheme, a surprisingly large influence of the high frequency modes is visible, which 
seems to be an artifact of the ad-hoc inclusion of the temperature correction, as it is not present in the (non-thermalized) NISE. Incorporating the thermal correction into the off-diagonal elements of the non-adiabatic coupling $S(t)$, as shown in Eq.~\ref{eq:JansenFactor_revisited}, makes the rate of thermalization dependent on $S(t)$. The magnitude of the off-diagonal elements of $S(t)$ increase with larger changes in the Hamiltonian between two time steps. The change in the Hamiltonian increases with every added bath mode, irrespective of whether these modes are in regions with relevant frequencies. It  seems  that this fact is the primary cause for the impact of high-frequency modes on TNISE populations, but needs to be investigated further. In previous publications on the  TNISE and MLNISE approaches, Drude spectral densities were employed, lacking  high-frequency components  that are common in realistic systems.  Thus, the present issue has not been observed in those studies \cite{jans18a,holt23a}. 

In conclusion, the introduced noise algorithm opens up possibilities for comparing population dynamics with different spectral densities. However, caution is advised when utilizing TNISE with spectral densities including high-frequency peaks, as it can introduce nonphysical thermalization effects. Potential solutions  include training MLNISE models on systems with high-frequency noise components, or utilizing a low-pass filter to specifically eliminate high-frequency components that should only minimally affect the population dynamics.
\section{Absorption spectra}
\label{sec:absorption}
\begin{figure}[htb]
\captionsetup[subfloat]{position=top,singlelinecheck=false,justification=raggedright,labelformat=brace,font=bf}
\subfloat[][]{
\includegraphics[width=0.485\textwidth]{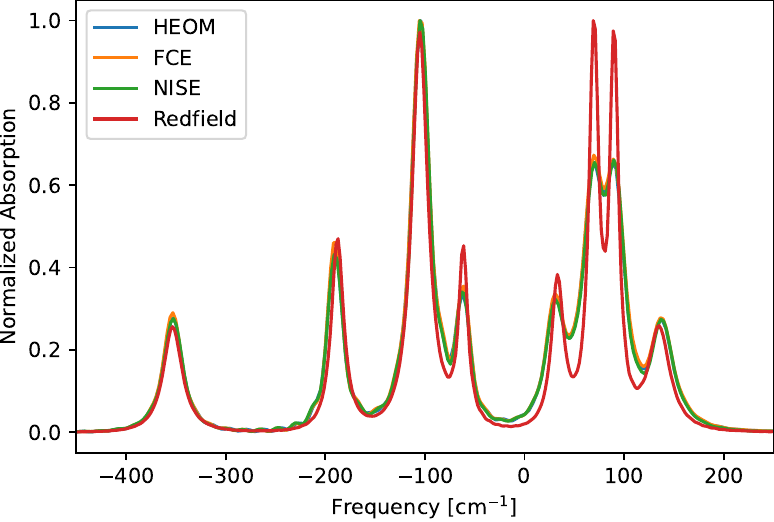}
}
\hskip0em\relax
\subfloat[][]{
\includegraphics[width=0.485\textwidth]{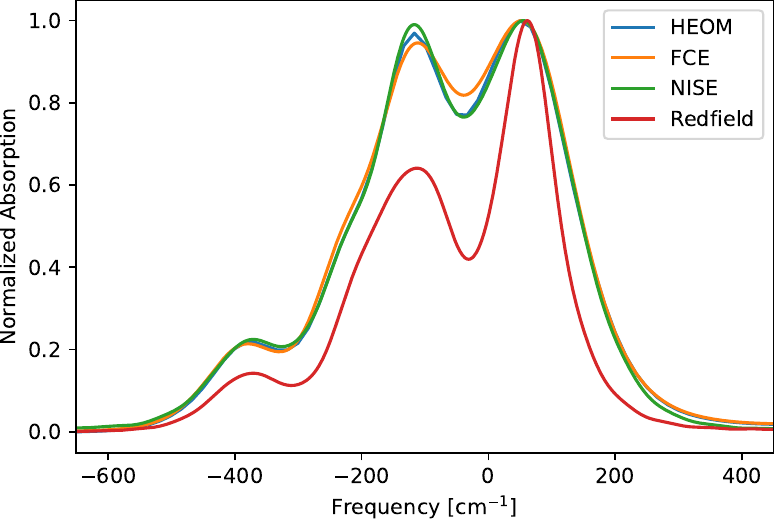}
}
\\[-4ex]
\subfloat[][]{
\includegraphics[width=0.485\textwidth]{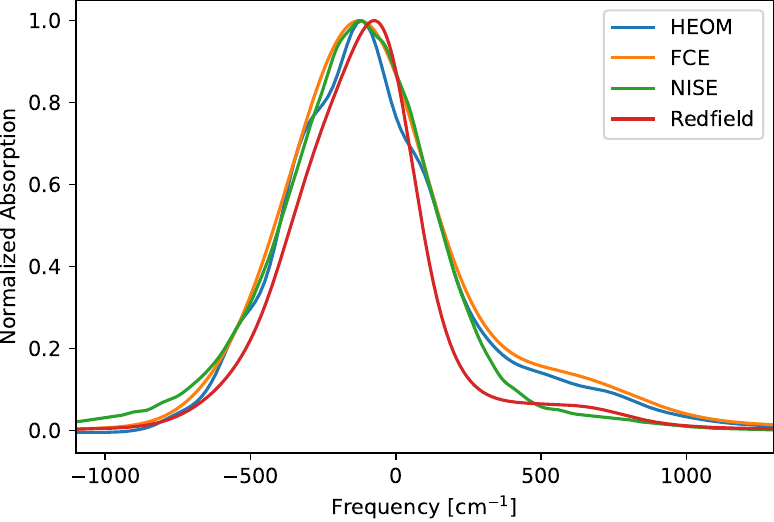}
}
\hskip0em\relax
\subfloat[][]{
\includegraphics[width=0.485\textwidth]{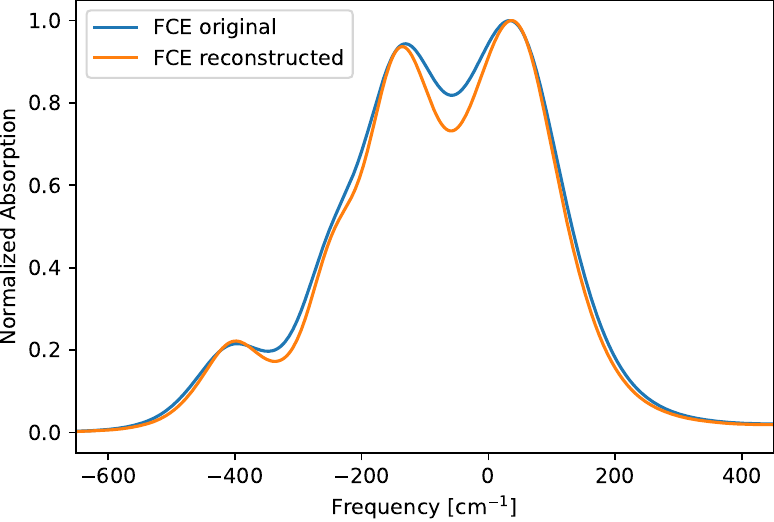}
}
\caption{Absorption spectra for the FMO complex utilizing the artificial spectral density as shown in Fig.~\ref{fig:high_freq_thermalization_issue}a with different reorganization energies and peaks:  a) three peaks with $\lambda = 6~\text{cm}^{-1}$,  b) two peaks with $\lambda =40~\text{cm}^{-1}$ and c) two peaks with $\lambda = 200~\text{cm}^{-1}$. The zero point for the frequency in the HEOM result was set to the average site energy of the Hamiltonian, and the spectra of the other methods have been shifted along the frequency axis to maximize the overlap with the HEOM result. Note that for panel a, the blue HEOM line and orange FCE line almost exactly match the green NISE line and are, therefore, barely visible.
To get a feeling to which degree uncertainties in the spectral density can affect an absorption spectrum,  in panel d, a comparison of the FCE-based spectra is shown for the original spectral density with three peaks and $\lambda =60~\text{cm}^{-1}$ alongside its reconstruction from a  
 40~ps-long noise trajectory. }
\label{fig:absorption}
\end{figure}

While population dynamics can often offer an understanding of excitonic pathways and give crucial insights into the inner workings of photosynthetic complexes, they are hard to verify experimentally. Conversely, one of the easiest quantities to obtain experimentally are absorption spectra. Hence, calculating absorption spectra with computational methods can be a first bridge between simulation and experiment. The previously discussed noise generating algorithm enables this comparison for arbitrary experimental spectral densities utilizing the NISE formalism. In this section, we present a comparative analysis with other popular methods such as the  Redfield-like approaches  \cite{novo10a,reng13a,juri15b} and the Full Cumulant Expansion (FCE) scheme \cite{cupe20b} using the HEOM formalism as benchmark. While more extensive analyses of different methods for calculating absorption spectra have certainly been performed, for example, by Zuehlsdorff et al.~\cite{zueh19a}, we aim at comparing absorption calculations based on the NISE scheme with other methods popular in the field of light-harvesting systems. For the FCE formalism, we utilized the implementation by Cupellini et al.~\cite{cupe20c} and for the Redfield theory we employed an in-house code utilized already in previous studies \cite{mait21b,sarn22a}. As was the case in the previous section, utilizing the HEOM as the benchmark prevents the use of very complicated spectral densities and, therefore, the three peak spectral density as shown in Fig.~\ref{fig:high_freq_thermalization_issue}a has been used for testing purposes. 

The time-domain absorption $\sigma (t)$ in the NISE framework can be derived from the time evolution matrix $U_{m,n}$ and the dipole moments $d_{0,n}$ of the individual pigments \cite{jans06a} (see also Appendix \ref{app:absorption})
\begin{equation}
  \sigma (t) =  \sum_{m,n}  d_{0,m}  U_{m,n}(t,0) d_{0,n}~.
\end{equation}
Versions that capture non-Condon effects, i.e., allowing time dependent dipole moments, have also been derived and are available, e.g., in the NISE 2017 code \cite{jans17a}.
The absorption spectrum can, subsequently, be obtained by taking the real part of the Fourier transform of $\sigma (t)$. Notably, the time-domain function $\sigma (t)$ usually rapidly decays to zero before thermal effects become significant. Hence, often barely any differences between the thermalized versions (TNISE and MLNISE) and the non-thermalized NISE approach are visible in the case of absorption spectra \cite{jans21a}. Therefore, the (non-thermalized) NISE is utilized for the following comparison. Figure \ref{fig:absorption} illustrates the absorption spectra obtained using the same FMO Hamiltonian as above, as well as dipole moments from Ref.~\cite{mait20a} together with the aforementioned three peak spectral density. One modification was done to the Hamiltonian as the GPU-HEOM tool \cite{krei13b} only supports up to 8 sites:  BChl 8a was dropped from the Hamiltonian of FMO, as it is coupled least strongly to the other pigments. The test was performed for three different reorganization energies $\lambda=6~\text{cm}^{-1}$, $40~\text{cm}^{-1}$ and $200~\text{cm}^{-1}$ obtained by scaling the spectral density respectively. The GPU-HEOM tool was only able to calculate absorption spectra for an 8 site Hamiltonian and 3 peaks up to a hierarchy depth of 4. At this depth, only the low reorganization energy case of $\lambda=6~\text{cm}^{-1}$ had converged. Therefore, the cases with $\lambda=40~\text{cm}^{-1}$ and $200~\text{cm}^{-1}$ were calculated using only the first two peaks. For two peaks, the GPU-HEOM tool was able to calculate up to a depth of 6 which was sufficient for the absorption spectra to converge.
A temperature of 300K was chosen for all calculations of absorption spectra here. Moreover, the parameters for the different methods 
were chosen, such that the absorption spectra converged visually. The simulation time was set to 4~ps for the Redfield method in the $\lambda=6~\text{cm}^{-1}$ case and 2~ps for all other methods. In the cases with $\lambda=40~\text{cm}^{-1}$ and $200~\text{cm}^{-1}$, a 0.5~ps simulation time was sufficient for all methods, while the time step was set to 1~fs in all cases. For the NISE approach, 50,000 realizations were used to show very smooth curves.
The heights of the  spectra were normalized to unity and along the frequency axis the spectra were aligned  to achieve  maximum overlap within the shown frequency range with the respective HEOM result. This is done because, in practice, absorption spectra from computational methods are usually shifted to match experimental spectra. Hence, a shift introduced by a computational method would not lead to worse results in practice. The unshifted spectra and calculated shifts can be found in Appendix~\ref{app:unshifted_Abs}. For the HEOM result, the zero point of the frequency axis was set to the average site energy of the Hamiltonian. 
As can be seen in Fig.~\ref{fig:absorption}a, for the reorganization energy of $\lambda=6~\text{cm}^{-1}$, the absorption spectra obtained from FCE, Redfield, and NISE approaches match the HEOM results quite well, demonstrating the robustness of these methods at low reorganization energies. The largest deviations can be seen for the Redfield approximation, while the FCE and NISE schemes almost perfectly agrees with the HEOM results. However, as the reorganization energy increases to $\lambda=40~\text{cm}^{-1}$, the accuracy of the Redfield approach deteriorates, highlighting its limitations in handling higher reorganization energies. At the same time, the NISE and FCE methods maintain a reasonable accuracy with a slightly improved agreement for the NISE procedure in this case, as can be seen in Fig.~\ref{fig:absorption}b.
In the case of $\lambda=200~\text{cm}^{-1}$ and as shown in Fig.~\ref{fig:absorption}c, the performance of both the NISE and the FCE schemes also degrades slightly. Notably, the NISE procedure fails to capture the vibrational sideband present in the spectra, which the FCE formalism even slightly overestimates the side band. This discrepancy underscores the challenges in accurately modeling systems with higher reorganization energies using the NISE scheme, especially when vibrational contributions become relevant. The comparisons up to this point have been done using the analytical form of the three-peak spectral density. As discussed above, it is also of interest how the absence of an analytically known spectral density affects the results. To this end, we generated a 40~ps-long noise trajectory (the same trajectory length utilized, e.g.,  in Ref.~\cite{mait20a}) trying to mimic the three-peak spectral density with $\lambda=60~\text{cm}^{-1}$. Subsequently, this trajectory was employed to re-calculate  the spectral density according to the procedure detailed in Section~\ref{sec:SD-reconstruction}. The difference in the absorption spectra for the FCE approach between the results using the analytical and the reconstructed spectral density is quite small, as delineated in Fig.~\ref{fig:absorption}d. Overall, the NISE method is remarkably accurate for its simplicity. This accuracy can be attributed to the aforementioned rapid decay of the time-domain function $\sigma(t)$ before thermal effects become significant. Besides wrong thermalization, the other major limitation is the assumption of a weak system-bath coupling, which explains the accuracy for low to medium reorganization energies. For larger reorganization energies, the FCE method stands out as it compares best to the HEOM benchmark results, even though its accuracy also starts deteriorating for larger reorganization energies. We note in passing that we did not test the modification of the Redfield theory here, in which the sites are first separated into strongly coupled subsets for which the absorption is being determined and then joined to a complete absorption spectrum \cite{mueh12b}.

\section{\label{sec:Conclusions} Conclusions}

The present study examined several aspects of  describing environmental effects on quantum dynamics, focusing particularly on the determination of spectral densities from fluctuating site energies and vice versa as well as on the
Numerical Integration of the Schrödinger Equation (NISE) approach including variants thereof. We specifically investigated interactions with classical baths and highly structured spectral densities. Our research aimed to refine these methodologies, addressing specific issues that hinder their accuracy and practicality, and establishing best practices.

In addressing the challenges within the NISE framework, a primary issue we tackled was the previously observed inaccurate representation of equilibrium distributions in thermalized versions of the NISE scheme \cite{holt23a}, namely the Thermalized NISE (TNISE) \cite{jans18a} and Machine Learned NISE (MLNISE) approaches \cite{holt23a}. To address this issue, we proposed an alternative ensemble-averaging procedure within the NISE framework. Our results demonstrate that this modified method significantly improves the long-time population dynamics achieved by both the TNISE and MLNISE schemes, thereby enhancing the accuracy of these quantum dynamic simulations.

Furthermore, we discussed a noise generation algorithm that, while not new \cite{mast87a,rava00a}, has been adapted from other fields to enable the use of the NISE approach with noise corresponding to (very) complicated spectral densities, which enables comparisons between NISE simulations done with MD or QM/MM noise and NISE simulations with noise following experimental spectral densities. Our investigations revealed, however, that employing TNISE with arbitrary spectral densities can introduce a too rapid thermalization, particularly when high-frequency components are present. Reference simulations using the Hierarchical Equations of Motion (HEOM) and the traditional NISE scheme showed no such effects, confirming that this issue is specific to the TNISE modification. 
The models for MLNISE, previously trained on systems with simple Drude spectral densities \cite{holt23a}, are not suited for systems with arbitrary spectral densities.
Training new models on systems with spectral densities that include also higher frequency components could enable the MLNISE scheme to handle systems with arbitrary spectral densities. Such a model should ideally learn to avoid the too rapid thermalization observed in the TNISE variant, presenting an exciting direction for future research.

The capability to generate noise corresponding to highly structured spectral densities has also facilitated the development 
of an optimized procedure (“best practice”) how to most efficiently extract complicated spectral densities in cases of a limited amount of data, i.e., site energy fluctuations. This result will be very useful in the future to reduce the numerical effort employed to accurately determine spectral densities and has been summarized, along with best practices for utilizing the NISE scheme, in Appendix~\ref{app:cookbooks}.

Additionally, we explored the accuracy of the NISE approach to calculate absorption spectra and found that, for such a relatively straightforward method, it performs remarkably well. In our tests with the FMO complex, the NISE procedure nearly matched the performance of Full Cumulant Expansion (FCE) and outperformed the  Redfield method, commonly used for absorption spectra of multi-chromophoric systems.

Having tested the thermalized version of the NISE approach even further, opens several avenues of applying the NISE or TNISE schemes in large to very large systems, such as chlorosomes \cite{vall14a,eric23a,eric23b,eric24a}, where the NISE has only been applied with energy fluctuations following Coulomb intermolecular interaction, rather than a defined spectral density. Another very large system of interest for the application of NISE or TNISE schemes are chromatophores \cite{sing19a}, as they host the entire network of bioenergetic proteins involved in converting absorbed sunlight to chemical energy stored in ATP. 

\section*{Acknowledgements} We are grateful to Marco David for initial
discussions on how to generate structured noise. This work has been supported
by grant KL 1299/24-1 and by the Research Training Group 2247 “Quantum Mechanical Materials Modeling” of the Deutsche Forschungsgemeinschaft (DFG).  Furthermore, the simulations were performed on a compute cluster funded through project INST 676/7-1 FUGG.

\section*{Data Availability} 
The data that supports the findings are available from the corresponding authors upon reasonable request. The implementation of the described algorithms is available at https://github.com/CPBPG

\appendix 

\section{\label{app:Cosine_Transform} Generating Spectral Densities from the Bath Correlation Function}

The spectral density  $J(\omega)$, which described the interaction of the primary system with the thermal bath,  is related to the cosine transform of the bath correlation function  $C(t)$ by
\begin{equation}
J(\omega) = \frac{\beta \omega}{\pi} \int_{0}^{\infty} C(t) \cos(\omega t) \, dt~.
\label{eq:jw_appendix}
\end{equation}
Similarly the bath correlation function $C(t)$ can also be obtained from the spectral density by utilizing the inverse cosine transform 
\begin{equation}
C(t) = \frac{ 2 }{ \beta} \int_{0}^{\infty} \frac{J(\omega)}{\omega} \cos(\omega t) \, d\omega~.
\label{eq:jw_appendix_inverted}
\end{equation}
In discrete terms, these relations can be calculated using a Discrete Cosine Transform (DCT) or Inverse DCT (IDCT) respectively. In the present case, we utilize a DCT-I, i.e., a  DCT of type one.
For real, even-symmetrical inputs, the type one DCT is equivalent to a Fast Fourier Transform (FFT) \cite{frig05a}. This equivalence allows us to compute the spectral density using the FFT by extending the bath correlation function $C(t)$  to form a symmetrical sequence. This is achieved by appending a reversed version of $ C(t)$, excluding its first and last elements, to itself. The FFT is then applied to this extended, symmetrical sequence, denoted as $C(t)_\text{ext,sym}$.

The spectral density is obtained from the real part of the FFT output, scaled appropriately to account for the physical constants and parameters of the system. The final expression for the spectral density using this FFT-based approach is given by
\begin{equation}
J(\omega) = \frac{\omega \Delta t}{ 2 \pi k_B T} \cdot \Re \{\textbf{FFT}(C(t)_\text{ext,sym})\}~,
\end{equation}
where $ \Delta t $ denotes the time step, $k_B$  the Boltzmann constant, and $T$  the temperature of the system.
Similarly, the bath correlation function can be obtained in a discrete way as
\begin{equation}
C(t) = \frac{2 \pi k T }{ \Delta t } \cdot \Re \left\{\textbf{IFFT}\left(\left(\frac{J(\omega)}{\omega}\right)_\text{ext,sym}\right)\right\}~.
\end{equation}
The definitions of $\textbf{FFT}$ and $\textbf{IFFT}$ used here correspond to those also employed in the Python library NumPy.

\section{\label{app:timeseries_resampling} Resampling Time Series}

The process of resampling a time series to a higher resolution involves increasing the number of data points and thereby reducing the time step between successive samples. This can be achieved in the frequency domain using a Fourier transformation, followed by zero-padding and an inverse Fourier transformation. For this purpose, any common signal processing library, such as \emph{scipy.signal} \cite{virt20a} can be used, but we will briefly describe the background here.

Consider  trajectories $E[t_i]$ of energy values with a time step $\Delta t_{\text{orig}}$. The goal is to obtain a new resampled time series $E_{\text{resamp}}[t_j]$ with a reduced time step $\Delta t_{\text{target}}$, where $\Delta t_{\text{target}} < \Delta t_{\text{original}}$.
The first step of tis process involves transforming $E[t_i]$ into the frequency domain using an FFT 
\begin{equation}
E[f_k] = \textbf{FFT}\{E[t_i]\} = \sum_{i=0}^{N-1} E[t_i] e^{-\frac{2\pi k i}{N}}~,
\end{equation}
where $N$ denotes the number of points in the original time series and $f_k$  the resepctive frequency with index $k$.
To increase the resolution in the time domain, we introduce additional points in the frequency domain by appending zeros. This process is known as zero-padding. The number of zeros to add, $Z$, is determined by the desired increase in resolution
\begin{equation}
Z = N \left(\frac{\Delta t_{\text{orig}}}{\Delta t_{\text{target}}} - 1\right).
\end{equation}
Zero-padding is applied symmetrically to the frequency spectrum $E[f_k]$ for the signal to stay real-valued after the inverse Fourier transformation
\begin{equation}
E_{\text{pad}}[f_k] = \begin{cases} 
E[f_k] & \text{for } k \leq \frac{N}{2} \\
0 & \text{for } \frac{N}{2} < k \leq \frac{N}{2} + Z \\
E[f_{k-Z}] & \text{for } k > \frac{N}{2} + Z
\end{cases}
\end{equation}
Finally, the inverse FFT (IFFT) is applied to the zero-padded frequency spectrum to convert it back into the time domain yielding the resampled time series
\begin{equation}
E_{\text{resamp}}[t_j] = \textbf{IFFT}\{E_{\text{pad}}[f_k]\} = \frac{1}{N+Z} \sum_{k=0}^{N+Z-1} E_{\text{pad}}[f_k] e^{\frac{2\pi k j}{N+Z}}~.
\end{equation}
This results in a new time series with a higher number of samples and a reduced time step, achieving the desired resampling effect. Such a simple resampling method only works without loss of information if frequencies above the Nyquist frequency are completely absent. If higher frequencies are present in the noise, whether due to artifacts or real signals, aliasing artifacts can occur. In such cases, a tapering window may be employed in the frequency domain to prevent artifacts introduced by a sharp transition to the padded region. Such a procedure does not prevent aliasing artifacts, i.e., the signal from the undersampled frequencies will still be wrongfully attributed to other frequencies. However, it removes “ringing”, i.e., oscillations caused by the sharp transition. If these higher frequencies become too strong, it makes sense to simply choose a higher sampling rate in the first place.

\section{\label{app:absorption} Absorption NISE}

The absorption spectrum is the Fourier transform of the autocorrelation of the dipole operator. In the interaction picture
with $\bar{\mu}_I(t) = \mathbf{U^{\dag}}(t,0)  \bar{\mu} \mathbf{U}(t,0)$, this is given by \cite{tokm14a}
\begin{equation}
  \sigma (\omega) =  \frac{1}{2 \pi} \int_{-\infty}^{+\infty} dt e^{-i \omega t} \langle\bar{\mu}_I(t)\bar{\mu}_I(0) \rangle \label{eq:1} ~,
\end{equation}
but for simplicity we will stay in  time domain in the following. For a proper thermal averaging, 
this expression needs to be evaluated over the trace of all possible initial eigenstates weighted by their thermal equilibrium probability \cite{tokm14a}
\begin{equation}
  \sigma (t) = \langle\bar{\mu}_I(t)\bar{\mu}_I(0) \rangle = \sum_n p_n \bra{n} \bar{\mu}_I(t)\bar{\mu}_I(0) \ket{n} ~.
\end{equation}
The Frenkel Hamiltonian considered in this study is confined to the single exciton manifold with no explicit inclusion of the ground state. The absorption formula, however, requires the ground state to be included. Since the ground state is much lower in energy compared to the single exciton manifold, the thermal equilibrium probability of finding the system in the ground state is effectively unity.
Thus, the equation simplifies to
\begin{equation}
  \sigma (t) =   \bra{0} \bar{\mu}_I(t)\bar{\mu}_I(0) \ket{0} ~.
\end{equation}
For the NISE approach, we need to transform back into the Schrödinger picture to obtain
\begin{equation}
  \sigma (t) =   \bra{0} \mathbf{U^{\dag}}(t,0)  \bar{\mu} \mathbf{U}(t,0)   \bar{\mu} \ket{0} ~.
\end{equation}
Inserting the identity operator in site bases three times yields  
\begin{equation}
\label{eq:absorption5}
  \sigma (t) =  \sum_{l,m,n} \bra{0} \mathbf{U^{\dag}}(t,0) \ket{l} \bra{l}   \bar{\mu} \ket{m} \bra{m}  \mathbf{U}(t,0) \ket{n} \bra{n}   \bar{\mu} \ket{0} ~.
\end{equation}
Due to the lack of coupling between the ground and excited states, the following relation holds $\bra{0} \mathbf{U^{\dag}}(t,0) \ket{l}= \mathbf{U}_{l,0}(t,0) =\delta_{0,l} \exp(iE_0t/ \hbar)$. Applying a Fourier transform,  $\exp(iE_0t/ \hbar)$ corresponds to a shift of the absorption spectrum along the frequency axis. Since the absorption spectrum is usually shifted to match the experimental spectrum anyway, this term can be dropped, leading to the final result
\begin{equation}
  \sigma (t) =  \sum_{m,n} \bra{0} \bar{\mu} \ket{m} \bra{m}  \mathbf{U}(t,0) \ket{n} \bra{n}   \bar{\mu} \ket{0}= \sum_{m,n}  d_{0,m}  U_{m,n}(t,0) d_{0,n}~.
\end{equation}
This expression can subsequently be Fourier transformed to obtain the absorption spectrum.
If no shifting is performed to match other results, one can instead shift the resulting spectrum by the frequency corresponding to the ground state energy $E_0 / \hbar$, which was previously dropped from Eq.~\ref{eq:absorption5}.

\section{\label{app:unshifted_Abs} Unshifted Absorption Spectra}

\begin{figure}[htb]
\captionsetup[subfloat]{position=top,singlelinecheck=false,justification=raggedright,labelformat=brace,font=bf}
\subfloat[][]{
\includegraphics[width=0.485\textwidth]{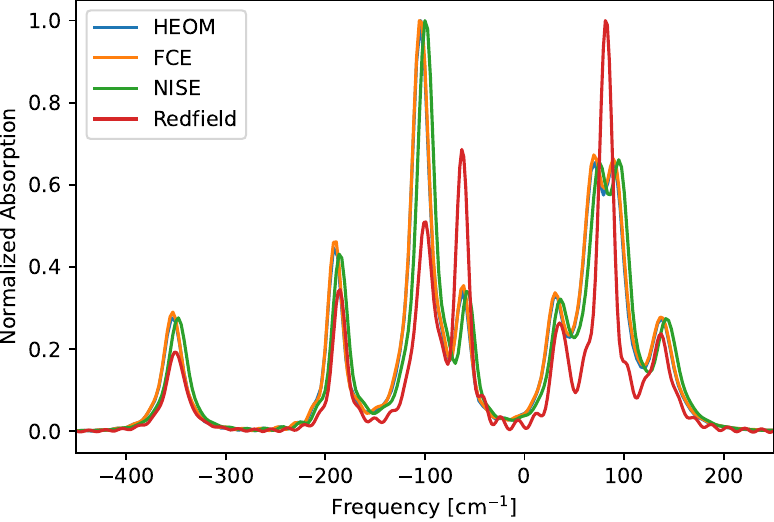}
}
\hskip0em\relax
\subfloat[][]{
\includegraphics[width=0.485\textwidth]{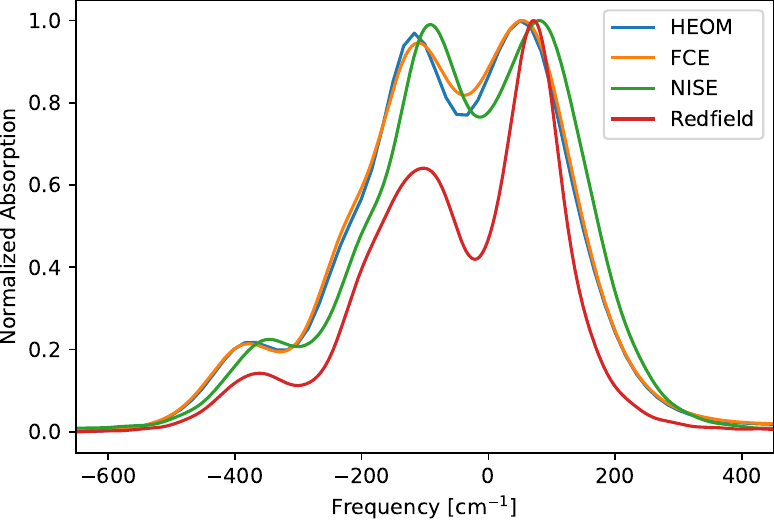}
}
\\
\subfloat[][]{
\includegraphics[width=0.485\textwidth]{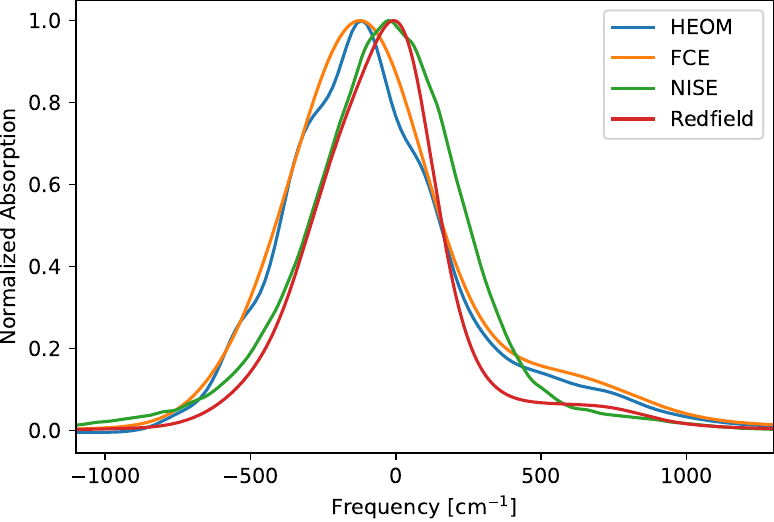}
}

\caption{Same as Fig. \ref{fig:absorption} but without shifting the results in energy to match the HEOM peak position.}
\label{fig:absorption_unshifted}
\end{figure}

In section \ref{sec:absorption} the absorption spectra were shifted in energy to match the HEOM peak positions. These shifts were justified as being common practice when comparing to experimental absorption spectra. However, from a theoretical perspective, it is worth noting such shifts, because it is an error introduced by the method. Therefore, Fig.~\ref{fig:absorption_unshifted} shows the absorption spectra without the shifts and Table~\ref{table:shifts} reports the applied shifts.

\begin{table}[!bt]
\begin{small}
\begin{center}
 \begin{tabular}{c c c c c c} 
 \hline
 \hline
  $\lambda$[cm$^{-1}$] &   shift FCE [cm$^{-1}$] & shift NISE [cm$^{-1}$] & shift Redfield [cm$^{-1}$]  \\ [0.5ex] 
 \hline\hline
 6 & -0.07 & -5.18 & -2.24 \\
  \hline
 40 & -0.88 & -25.19 & -9.79 \\
   \hline
 200 & -2.88 & -97.93 & -66.25 &  \\
   \hline
 \hline\hline
\end{tabular}
\end{center}
\end{small}
 \caption{shifts applied to absorption spectra in Fig.~\ref{fig:absorption} compared to the unshifted spectra in Fig.~\ref{fig:absorption_unshifted}}
 \label{table:shifts}
\end{table}

\section{\label{app:super_resolution} Summary of Super-Resolution Method and Modifications}
To implement super-resolution technique, we mostly follow the procedure outlined in Ref.~\onlinecite{mark16a}.
The basic idea is to approximate the auto-correlation $C(t)$ with a series of Drude-Lorentz terms
\begin{equation}
C(t) = \sum_{i,j} \lambda_{ij} e^{-\gamma_i t} \cos(\Omega_j t),
\end{equation}
where $\lambda_{ij}$ are the expansion coefficients, $\gamma_i$ the linewidths, and $\omega_j$  the respective frequencies.
This expression can be rewritten in matrix form as a linear algebra problem
\begin{equation}
C_k = A_{ijk} \lambda_{ij},
\label{eq:super:linalg}
\end{equation}
where $C_k = C(t_k)$ represents the discretized correlation function at different time points and $A_{ijk} = e^{-\gamma_i t_k} \cos(\Omega_j t_k)$ denotes the coefficient matrix. Here, the indices $i, j,$ and $k$ correspond to the grid points for damping coefficients, frequencies, and time, respectively.
To ensure the system is underdetermined and to allow for the selection of a suitable solution, the grid of $\gamma_i$ and $\Omega_j$ is chosen such that there are significantly more basis functions than time samples. This underdetermination enables us to “choose” a suitable solution by introducing regularization.
However, for the complicated FMO spectral density studied in this work, the total variation norm proposed for regularization in Ref.\onlinecite{mark16a} did not perform well.  It resulted in many unphysical, negative peaks in the spectral density. To address this issue, we replaced the total variation norm with a penalty for negative peaks.
Specifically, we rephrased the problem as a minimization of the following objective function
\begin{equation}
a \|A_{ijk} \lambda_{ij} - C_k\|_2 + b \|\lambda_{ij}\|_1 + c \sum_{ij} |\lambda_{ij}| -  \lambda_{ij},
\label{eq:super:objective}
\end{equation}
where  $\sum_{ij} |\lambda_{ij}| -  \lambda_{ij}$ is the negative penalty. The parameters $a, b,$ and $c$ need to be chosen to balance the terms in the optimization. For the calculations in this work $a = 1 \cdot 10^4$, $b = 1$, and $c = 0.1$ were used.
Instead of using the two-step iterative shrinkage tresholding (TwIST) \cite{biou07a} algorithm as proposed in Ref.~\onlinecite{mark16a}, which requires determining the proximal operator for the negative penalty, we opted for the limited memory Broyden–Fletcher–Goldfarb–Shanno (L-BFGS) optimization \cite{liu89b}, specifically the  implementation in the PyTorch library \cite{pasz19a} with GPU acceleration. This choice allows for more straightforward handling of the penalty without additional complexity.
For our calculations, we used a grid in frequency space $\Omega_j$
with a range from 0 to 0.2 eV and intervals of $5 \cdot 10^{-5}$ eV. As well as a range of linewidths $\gamma_i$ from 1 to 200 cm$^{-1}$ with intervals of 0.5 cm$^{-1}$.
These very fine-grained grids were necessary to achieve good performance with our reference spectral density, but also led to very high computational costs. 
After the optimization, a debiasing step is performed. Only the top contributing modes are selected and a least squares optimization is performed with just those modes. In the present case we modified the procedure again and used a non-negative least squared optimization to avoid the negative peaks. Furthermore, we could not obtain suitable solutions with just a few peaks, so that we had to set the threshold for included peaks to a very low value of $5 \cdot 10^{-8}$.
The spectral density is then reconstructed from the determined parameters as
\begin{equation}
J(\omega)= 2 \pi \omega \sum_{i,j} \left( \frac{\lambda_{i,j} \gamma_i}{\gamma_i^2 + (\omega +\Omega_j)^2} + \frac{\lambda_{i,j}  \gamma_i}{\gamma_i^2 + (\omega -\Omega_j)^2} \right)
\end{equation}
\begin{figure}
\captionsetup[subfloat]{position=top,singlelinecheck=false,justification=raggedright,labelformat=brace,font=bf}
   \centering
   \subfloat[][]{
   \includegraphics[width=0.485\textwidth]{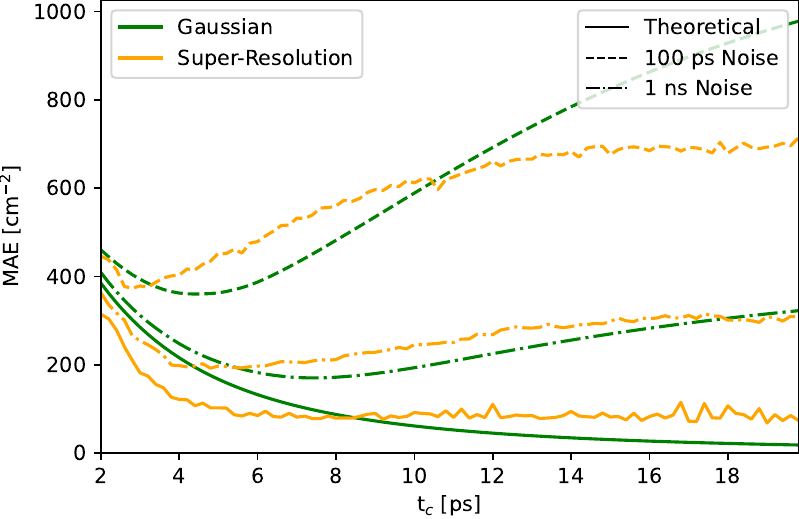}
    }
   \hskip0em\relax
   \subfloat[][]{
   \includegraphics[width=0.485\textwidth]{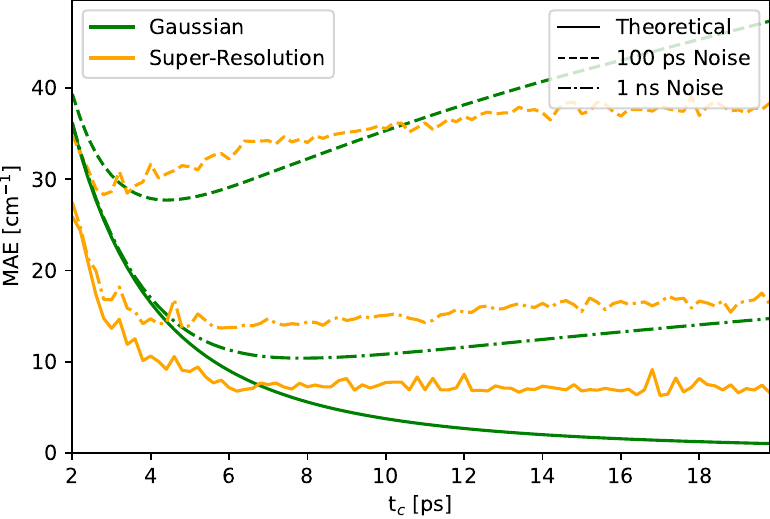}
    }
  \caption{Same  as in Fig.\ref{fig:errors} but only for the super-resolution technique and Gaussian damping.}
  \label{fig:errors:super}
\end{figure}
Fig.~\ref{fig:errors:super} shows the performance of the super-resolution technique compared to the Gaussian damping that was determined to be the overall best of the FFT methods in Sec.~\ref{sec:SD-reconstruction}. The cutoff time $t_c$ for the super-resolution technique corresponds to the length of the autocorrelation included in Eq.~\ref{eq:super:linalg}, which is a cutoff similar to the step function. In the theoretical case, the super-resolution scheme shows the best performance for low $t_c$, but stops improving after roughly $t_c=6$ ps. Increasing $a$ from the objective function in Eq.~\ref{eq:super:objective} would force solutions closer to the given autocorrelation, thereby improving the results for larger $t_c$ in the theoretical case but also the chance for overfitting in noisy cases. In the noisy cases, the super-resolution technique performs similarly but does not achieve as low a MAE as the Gaussian damping.
\begin{figure}
\captionsetup[subfloat]{position=top,singlelinecheck=false,justification=raggedright,labelformat=brace,font=bf}
   \centering
   \subfloat[][]{
   \includegraphics[width=0.485\textwidth]{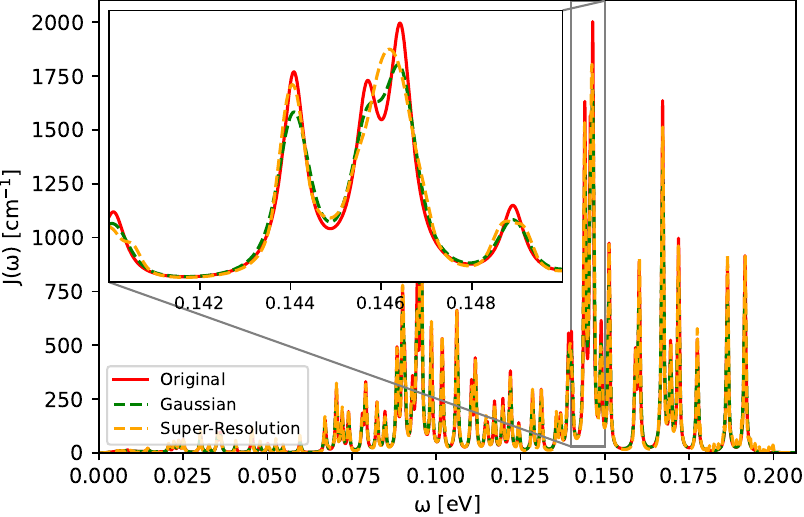}
    }
   \hskip0em\relax
   \subfloat[][]{
   \includegraphics[width=0.485\textwidth]{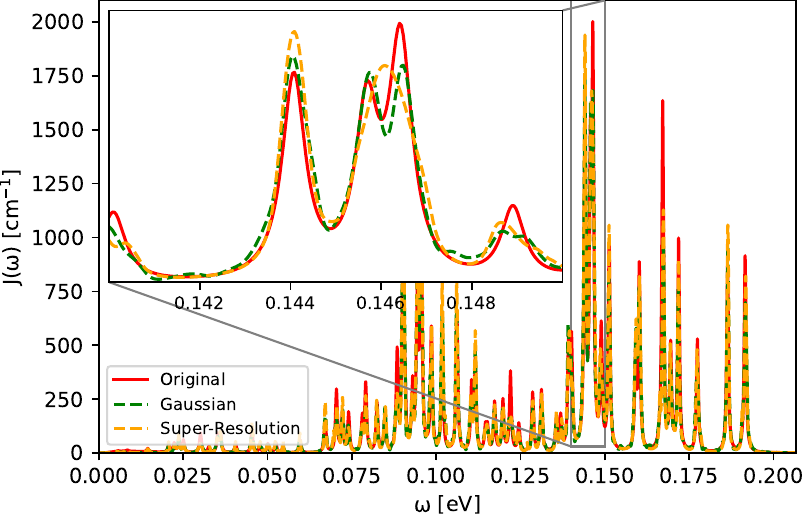}
    }
  \caption{Same as in Fig.\ref{fig:damping_Spectral_density} but  only for the super-resolution techniques and Gaussian damping. Cutoff times $t_c$ of 5 ps for the Gaussian damping and 3 ps for super-resolution were chosen based on the optimal performance displayed in Fig.~\ref{fig:errors:super}.}
  \label{fig:damping_Spectral_density:super}
\end{figure}
Similar to Fig.~\ref{fig:damping_Spectral_density} in Sec.~\ref{sec:SD-reconstruction}, Fig.~\ref{fig:damping_Spectral_density:super} contrasts the spectral densities determined by super-resolution and Gaussian damping with the original spectral density. Panel a shows spectral densities obtained from the theoretical autocorrelation function, while panel b for an autocorrelation functions based on a 100 ps noise trajectory. The cutoff times were chosen to be 5 ps for the Gaussian damping and 3 ps for super-resolution case based on the optimal performance displayed in Fig.~\ref{fig:errors:super}b for the 100 ps case. It can generally be seen that the Gaussian damping resolves details somewhat more accurate, while the super-resolution results align slightly better  in terms of peak heights. Overall, both methods offer very similar performance. However, the quality of the super-resolution results strongly depend  on a good choice of regularization on top of $t_c$. For example, the regularization proposed in Ref.~\onlinecite{mark16a} lead to completely unusable results for the reference spectral density considered here. In the present case, it was also necessary to consider a very fine-grained grid of line shapes and peak positions, making the calculation computationally demanding. While it is likely, that an appropriate regularization term could theoretically be found to make the super-resolution technique outperform Gaussian damping, doing so would be very challenging in any realistic scenarios where the correct spectral density is unknown. In most cases, the FFT-based damping approach will be preferable, as it is very easy to compute and only depends on one parameter, i.e.,  the cutoff time $t_c$. A very simple way for determining $t_c$ was described in Sec.~\ref{sec:SD-reconstruction}.

\section{\label{app:cookbooks} Recipes for Best Practices}

\subsection{Spectral densities from time series of site energies}

In this subsection, we summarize the results of section \ref{sec:SD-reconstruction} and present a step-by-step recipe for determining spectral density from limited data, such as excited state calculations along MD or QM/MM trajectories.
\begin{enumerate}
  \item Choose a time step as large as possible while still resolving all relevant frequencies according to \cite{nyqu28a,shan49a,unse00a}
  \begin{equation}
    \omega_{\text{nyq}}=2 \pi\frac{f_{\text{s}}}{2}=\frac{\pi}{\Delta t} .
\end{equation}
For chlorophyll molecules or similar, the relevant frequencies reach up to about 0.2 eV, which means a time step of 10 fs is sufficient. 
  \item For the length of the simulation, the longer, the better. The constraining factor is the computation time.
\item Calculate the autocorrelation function according to \cite{damj02a}
\begin{equation}
    C(t_j) = \frac{1}{N-j} \sum_{i=1}^{N-j} \Delta E(t_i + t_j) \Delta E(t_i)~.
\end{equation}
\item Find the time $t_c$ at which the autocorrelation function starts being dominated by noise. This can be done, for example, by looking at the moving absolute average (see Fig.~\ref{fig:autocorrelation}b) and does not have to be very accurate.
\item Apply a Gaussian damping function $\exp(-(t/t_c)^2)$ with cutoff time $t_c$ to the autocorrelation function \cite{loco18b}.
\item Perform  A Discrete Cosine Transform of C(t) and multiply the results by $\beta\omega/\pi$ to get the spectral density. See Appendix \ref{app:Cosine_Transform} on how to use an FFT instead.
\end{enumerate}

\subsection{NISE Variants in Combination with Arbitrary Spectral Densities}
\begin{enumerate}
  \item As a start, one needs to define a time-independent Hamiltonian of the system
  \begin{equation}
	\hat{H}_S=\sum_{i=1}^N E_i \ket{i} \bra{i} +\sum_{i=1}^N\sum_{j=1}^N
	V_{ij} \ket{i} \bra{j}.
\end{equation}  
  \item For every site $j$, generate a noise trajectory with the desired simulation length, that follows the desired spectral density $J_j(\omega)$. See next recipe for details.
  \item Propagate the wave function according to 
  \begin{equation}
\ket{\tilde{\psi}(t+\Delta t)}= \hat{S}(t) \Tilde{\hat{U}}(t+ \Delta t , t) \ket{\tilde{\psi}(t)}~,
\end{equation}
or choose $S^T$ or $S^{ML}$ for TNISE or MLNIE, respectively. (See Section \ref{sec:NISE} and \ref{subsec:thermal_correction_ensemble_averaged_dynamics} for more details.  Also, it is not advisable to use the MLNISE for arbitrary spectral densities without training a new model.)
\item Repeat for a number of realizations and average the results to get an ensemble average.
\item Be cautious when using TNISE concerning the early thermalization issue detailed in Section \ref{sec:population_dynamics}.
\end{enumerate}

\subsection{Generate Noise following Arbitrary Spectral Densities}
\begin{enumerate}
  \item Determine the target power spectrum of the desired spectral density $J(\omega)$ as
  \begin{equation}
    \tilde{C}^T(\omega)= \frac{2 \pi}{\beta \omega} J(\omega)
\end{equation}
\item Generate  white noise $\eta^W(t)$ with mean  zero and variance one for twice as many time steps as needed for the trajectory.
\item Perform a Fast Fourier Transform (FFT) of the white noise and divide it by the square root of the time step $dt$ to obtain the noise in frequency domain
\begin{equation}
    \tilde{\eta}^W(\omega)= \frac{\textbf{FFT}(\eta^W(t))}{\sqrt{dt}}~.
\end{equation}
\item Multiply the resulting noise in Fourier space with the square root of the target power spectrum to obtain the target noise $\tilde{\eta}^T(\omega)$ in Fourier space
\begin{equation}
    \tilde{\eta}^T(\omega)= \tilde{\eta}^W(\omega) \sqrt{ \tilde{C}^T(\omega)}~.
\end{equation}
\item Determine the real part of the inverse FFT resulting in the target noise $\eta^T(t)$
\begin{equation}
    \eta^T(t)= \Re (\textbf{IFFT}(\tilde{\eta}^T(\omega)) )~.
\end{equation}
\item Drop the second half of the noise to obtain noise of the desired length. This trick needs to be performed since the algorithm generates periodic trajectories otherwise. 
\end{enumerate}
\bibliography{ukleine.bib}

\end{document}